\documentclass[conference]{IEEEtran}

\usepackage{xspace}
\usepackage{url}
\usepackage{cite}
\usepackage{amsmath,amssymb,amsfonts,amsthm}
\usepackage[hidelinks]{hyperref}
\usepackage{enumitem} 
\usepackage{mdframed}
\usepackage{booktabs}
\usepackage{multirow}
\usepackage{algorithm}
\usepackage{graphicx}
\usepackage{esvect}

\newtheorem{definition}{Definition}

\usepackage{xcolor}
\usepackage{etoolbox}

\usepackage[noend]{algpseudocode}
\usepackage{mathtools} 

\setlist[itemize]{leftmargin=2em}
\setlist[enumerate]{leftmargin=2em}

\usepackage{tikz}
\usetikzlibrary{positioning, arrows.meta, backgrounds, calc, shapes.misc}
\usepackage{pgfplots}
\pgfplotsset{compat=1.18}

\usepackage{placeins}
\usepackage{colortbl}
\definecolor{RoyalBlue}{RGB}{65, 105, 225}
\definecolor{SteelBlue}{RGB}{70, 130, 180}
\definecolor{Orange}{RGB}{255, 165, 0}
\definecolor{Teal}{rgb}{0.0, 0.5, 0.5}
\definecolor{bluegray}{rgb}{0.4, 0.6, 0.8}
\definecolor{antiquebrass}{rgb}{0.8, 0.58, 0.46}
\definecolor{amethyst}{rgb}{0.6, 0.4, 0.8}
\definecolor{darkpastelgreen}{rgb}{0.01, 0.75, 0.24}
\definecolor{cbpurple}{RGB}{204, 121, 167}
\definecolor{cbgreen}{RGB}{0, 158, 115}

\renewcommand{\tablename}{Table}
\usepackage{caption}
\newcommand{\tool}{$\textsc{Beak}$\xspace}
\newcommand{\zkap}{$\textsc{ZKap}$\xspace}
\newcommand{\arguzz}{$\textsc{Arguzz}$\xspace}

\definecolor{hlcolor}{HTML}{156082}

\newcommand{\reducedstrut}{\vrule width 0pt height .1\ht\strutbox depth .1\dp\strutbox\relax}

\newcommand{\icode}[1]{%
  \begingroup
  \setlength{\fboxsep}{0pt}%
  \colorbox{gray!20}{\reducedstrut\textup{\texttt{\small{#1\xspace}}}\/}%
  \endgroup
}

\begin{document}
\sloppy

\title{Revealing Semantic Gaps in Zero-Knowledge Virtual Machines via Obligation-Targeted Semantic Fuzzing}

\author{\IEEEauthorblockN{Hongbo Wen\IEEEauthorrefmark{1},
Jingyu Ke\IEEEauthorrefmark{2},
Hanzhi Liu\IEEEauthorrefmark{1},
Yanju Chen\IEEEauthorrefmark{3}, and
Yu Feng\IEEEauthorrefmark{1}}
\IEEEauthorblockA{\IEEEauthorrefmark{1}University of California, Santa Barbara\\
\{hongbowen,hanzhi\}@ucsb.edu, yufeng@cs.ucsb.edu}
\IEEEauthorblockA{\IEEEauthorrefmark{2}Shanghai Jiao Tong University\\
windocotber@sjtu.edu.cn}
\IEEEauthorblockA{\IEEEauthorrefmark{3}University of California, San Diego\\
yanju@ucsd.edu}}

\hypersetup{
  pdftitle={Revealing Semantic Gaps in Zero-Knowledge Virtual Machines via Obligation-Targeted Semantic Fuzzing},
  pdfauthor={Hongbo Wen, Jingyu Ke, Hanzhi Liu, Yanju Chen, Yu Feng}
}

\IEEEoverridecommandlockouts
\makeatletter\def\@IEEEpubidpullup{6.5\baselineskip}\makeatother
\IEEEpubid{\parbox{\columnwidth}{
		Network and Distributed System Security (NDSS) Symposium 2027\\
		22--26 March 2027, Seoul, Republic of Korea\\
		ISBN 978-1-970672-09-1\\
		https://dx.doi.org/10.14722/ndss.2027.232037\\
		www.ndss-symposium.org
}
\hspace{\columnsep}\makebox[\columnwidth]{}}

\maketitle

\begin{abstract}
Zero-knowledge virtual machines (zkVMs) prove correct execution of programs and underpin systems securing billions of dollars in digital assets. Their security depends on alignment between the \emph{instruction set architecture} (ISA) that programs target, such as RISC-V, and the \emph{constraint system}, polynomial equations that check the prover's record of every intermediate value (the \emph{witness}). Mismatches, or \emph{semantic gaps}, come in two kinds: \emph{underconstraint} bugs let a malicious prover forge a proof of an execution the ISA forbids (a soundness violation), and \emph{divergence} bugs make the zkVM itself deviate from the ISA or crash, so an honest prover cannot prove a legitimate execution (a completeness violation). Both cluster at corner cases such as division by zero that conventional fuzzing almost never reaches, and underconstraint bugs leave program output unchanged. Our key insight is that these corner cases correspond to \emph{constraint obligations} derived from the ISA specification and the proof system's algebraic structure. We build \tool, a reinforcement-learning fuzzer in which obligations replace three random decisions: they supply inputs that trigger each obligation, reward executions that trigger new combinations of obligations, and direct witness tampering to the step where an obligation is triggered. \tool flags divergence bugs by comparing registers against a reference RISC-V emulator, and underconstraint bugs by checking whether the constraints accept a tampered witness. On six leading production RISC-V zkVMs, \tool detects 25 of 37 known bugs (67.6\%) versus 5 (13.5\%) for the prior state of the art, and discovers 14 previously unknown bugs, all responsibly disclosed and patched.
\end{abstract}

\IEEEpeerreviewmaketitle

\section{Introduction}\label{sec:intro}

A \emph{zero-knowledge proof} (ZKP) lets a prover convince a verifier that a computation produced a given output without revealing intermediate state, with a succinct proof the verifier checks far faster than recomputation~\cite{goldwasser}.  A \emph{zero-knowledge virtual machine} (zkVM) lifts this guarantee to general programs: instead of hand-writing a custom proof circuit (an arithmetic encoding of one specific computation) per application, the developer writes ordinary code targeting a standard \emph{instruction set architecture} (ISA), the fixed contract that defines what each machine instruction computes (most production systems target RISC-V), and the zkVM proves the entire execution~\cite{risc0,sp1,jolt,openvm,nexus,pico}.  zkVMs underpin blockchain rollups, privacy-preserving applications, and off-chain computation platforms collectively securing billions of dollars in digital assets; SP1-based deployments alone account for over \$5.5B in total value secured across 35+ customers~\cite{succinct_tvs}.  A bug in the equations a zkVM uses to check an execution (its \emph{constraint system}) lets a malicious prover forge proofs for executions that never happened. This is catastrophic in this setting, because the only artifact the attacker hands over is the proof itself, already accepted as valid by the verifier, so forgeries are invisible by construction.  The risk is concrete: a 2018 soundness flaw in Zcash's original proof system (BCTV14) could have allowed unbounded minting of its privacy-preserving (shielded) currency against a then-${\sim}\$1$B market capitalization before the Sapling upgrade patched it~\cite{zcash_counterfeit, zcash_market_cap_fortune}.

Internally, a zkVM's \emph{executor} runs the program and records an execution trace, which forms the \emph{witness} that the constraint system checks before a proof is generated (\autoref{sec:bg-zkvm}).  Bugs cluster where the executor and the constraint system must agree with the ISA; field arithmetic differs from 32-bit hardware in ways that force every operation to be reconstructed, and every reconstruction is a place where the encoding can silently diverge from the ISA.  We call any such deviation a \emph{semantic gap}, of two operationally distinct kinds: an \emph{underconstraint} bug, where the constraints accept an alternative witness that encodes an execution the ISA forbids (a soundness violation exploitable by a malicious prover); and a \emph{divergence} bug, where the executor itself deviates from the ISA on a legitimate input (e.g., a crash on a valid program), yielding a completeness violation.

Detecting these gaps is hard for two reasons.  First, the gaps cluster at non-trivial corner cases (division by zero, signed overflow, executions long enough to overflow a power-of-two-sized table) that conventional fuzzing almost never reaches.  Second, reaching a gap is not the same as seeing it.  When a test program hits a corner case, a divergence shows up as a wrong result or a crash, but an underconstraint shows nothing: the executor computes the right result and its witness satisfies the constraints, which are too permissive rather than wrong on this run.  Revealing it requires an alternative witness, one encoding an execution the ISA forbids, that the constraints still accept.  No program produces such a witness; it must be constructed deliberately from what the constraints are supposed to enforce, a kind of input that conventional fuzzing has no notion of.  Existing tools each address only part of the problem.  Static analyzers (e.g., \zkap~\cite{zkap}) target circuits written in domain-specific languages such as Circom and miss zkVM-level instruction semantics; symbolic tools (e.g., Picus~\cite{picus}) scale only to thousands of constraints, orders of magnitude below a production zkVM; prior dynamic testers (e.g., \arguzz~\cite{arguzz}) combine metamorphic testing with random fault injection and target both kinds of gap, but neither their input generation nor their fault injection is guided toward the corner cases where bugs hide.  A generic fuzzer guided by reinforcement learning (RL), which learns which mutations to apply from a reward signal, fails at three points: its seeds, typically standard test suites such as \icode{riscv-tests}~\cite{riscvtests}, exercise ISA-level edge cases but rarely the zkVM-specific ones, such as values crossing byte boundaries or executions that overflow a power-of-two-sized table; its reward, edge coverage of the executor's code, is poorly correlated with constraint structure; and its output comparison cannot see underconstraints, while blindly perturbing a witness of millions of values has negligible hit probability within any practical budget.

Our key insight is that these corner cases are not random: they correspond to \emph{constraint obligations} derived from two fixed sources: (i)~the ISA specification, which dictates what each instruction must compute, and (ii)~the algebraic structure of the proof system, which dictates how each computation must be encoded.  Applied to RV32IM, the 32-bit integer and multiplication subset of RISC-V, this two-source derivation yields a closed set of 60 obligations (\autoref{sec:trace-ir}) shared across six leading production RV32IM zkVMs.  We specify each obligation as a triple $(\mu_c, \kappa_c, \iota_c)$: an enrichment template $\mu_c$ that builds inputs triggering the obligation, an activation predicate $\kappa_c$ that recognizes when a trace step triggers it, and an injection template $\iota_c$ that injects an alternative witness at that step, derived once per obligation and reused across all backends.

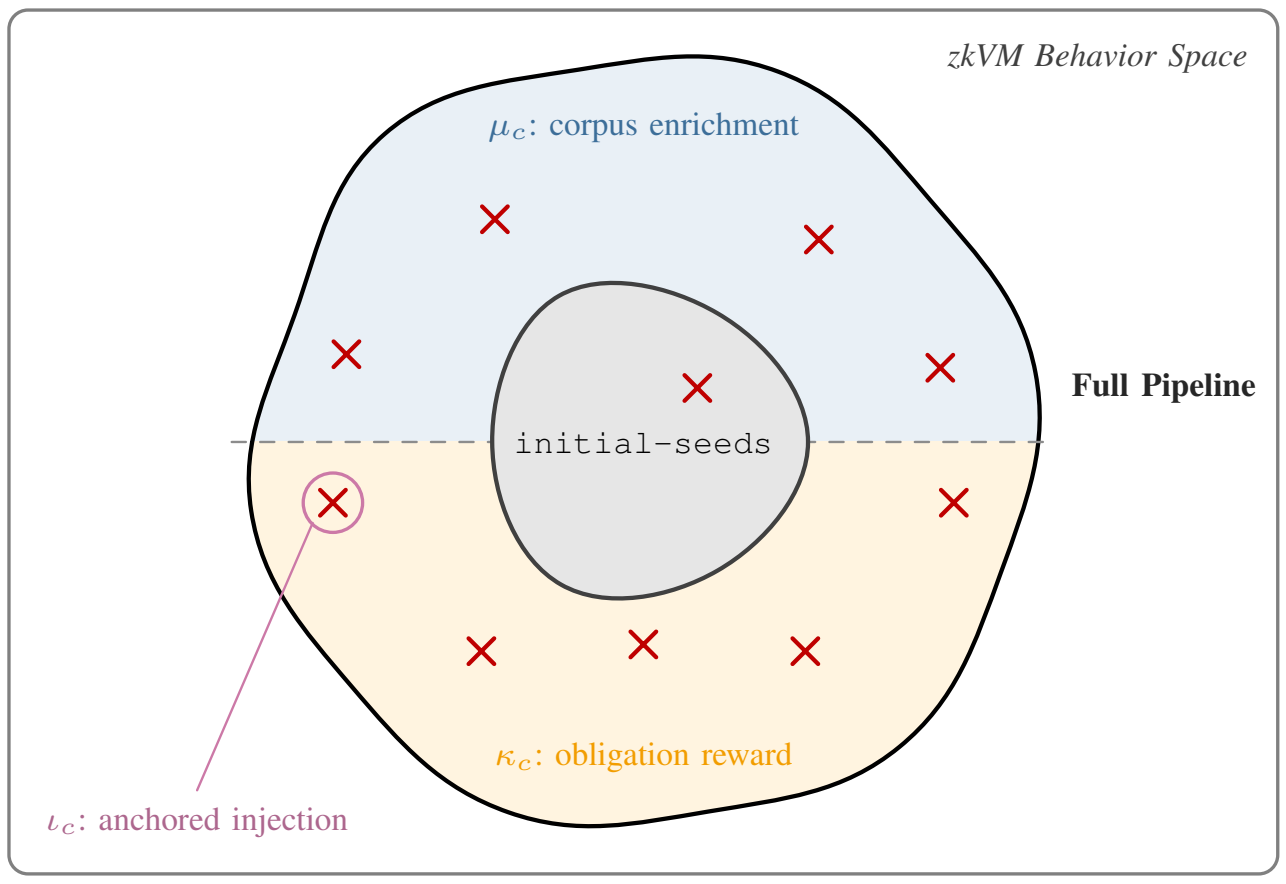
\begin{figure}[t]
    \centering
    \resizebox{0.95\columnwidth}{!}{%
    \begin{tikzpicture}[>=Stealth, line cap=round,
      bug/.style={cross out, draw=red!75!black, line width=0.9pt, minimum size=4.5pt, inner sep=0pt},
    ]
    \draw[draw=black!50, line width=0.7pt, rounded corners=4pt]
      (-4.7, -3.20) rectangle (4.7, 3.20);
    \node[font=\scriptsize\itshape, text=black!75, anchor=north east]
      at (4.60, 3.08) {zkVM Behavior Space};

    \begin{scope}
      \clip (-3.5, 0) rectangle (3.5, 3.5);
      \draw[draw=black, fill=SteelBlue!12, line width=0.9pt]
        plot[smooth cycle, tension=0.7] coordinates {
          ({2.95*cos( 10)}, {2.95*sin( 10)})
          ({2.78*cos( 40)}, {2.78*sin( 40)})
          ({2.92*cos( 70)}, {2.92*sin( 70)})
          ({2.80*cos(100)}, {2.80*sin(100)})
          ({2.98*cos(130)}, {2.98*sin(130)})
          ({2.75*cos(160)}, {2.75*sin(160)})
          ({2.95*cos(190)}, {2.95*sin(190)})
          ({2.82*cos(220)}, {2.82*sin(220)})
          ({2.92*cos(250)}, {2.92*sin(250)})
          ({2.78*cos(280)}, {2.78*sin(280)})
          ({2.95*cos(310)}, {2.95*sin(310)})
          ({2.82*cos(340)}, {2.82*sin(340)})
        };
    \end{scope}
    \begin{scope}
      \clip (-3.5, -3.5) rectangle (3.5, 0);
      \draw[draw=black, fill=Orange!12, line width=0.9pt]
        plot[smooth cycle, tension=0.7] coordinates {
          ({2.95*cos( 10)}, {2.95*sin( 10)})
          ({2.78*cos( 40)}, {2.78*sin( 40)})
          ({2.92*cos( 70)}, {2.92*sin( 70)})
          ({2.80*cos(100)}, {2.80*sin(100)})
          ({2.98*cos(130)}, {2.98*sin(130)})
          ({2.75*cos(160)}, {2.75*sin(160)})
          ({2.95*cos(190)}, {2.95*sin(190)})
          ({2.82*cos(220)}, {2.82*sin(220)})
          ({2.92*cos(250)}, {2.92*sin(250)})
          ({2.78*cos(280)}, {2.78*sin(280)})
          ({2.95*cos(310)}, {2.95*sin(310)})
          ({2.82*cos(340)}, {2.82*sin(340)})
        };
    \end{scope}

    \draw[draw=black!45, line width=0.5pt, dashed]
      (-3.05, 0) -- (3.05, 0);

    \draw[draw=black!75, fill=black!10, line width=0.9pt]
      plot[smooth cycle, tension=0.7] coordinates {
        ({1.22*cos(  0)}, {1.22*sin(  0)})
        ({1.10*cos( 60)}, {1.10*sin( 60)})
        ({1.25*cos(120)}, {1.25*sin(120)})
        ({1.12*cos(180)}, {1.12*sin(180)})
        ({1.22*cos(240)}, {1.22*sin(240)})
        ({1.10*cos(300)}, {1.10*sin(300)})
      };

    \node[bug] (innerbug) at ( 0.40,  0.40) {};

    \node[bug] at (-2.20,  0.65) {};
    \node[bug] at (-1.10,  1.65) {};
    \node[bug] at ( 1.30,  1.50) {};
    \node[bug] at ( 2.20,  0.55) {};

    \node[bug] at (-2.30, -0.45) {};
    \node[bug] at (-1.20, -1.55) {};
    \node[bug] at ( 0.00, -1.50) {};
    \node[bug] at ( 1.20, -1.55) {};
    \node[bug] at ( 2.30, -0.45) {};

    \draw[draw=cbpurple, line width=0.7pt]
      (-2.30, -0.45) circle (0.22);
    \node[font=\scriptsize, text=cbpurple!85!black, anchor=south west]
      (trglbl) at (-4.55, -3.05) {$\iota_c$:~anchored injection};
    \draw[draw=cbpurple, line width=0.5pt]
      (-2.45, -0.60) -- (trglbl.north);

    \node[font=\scriptsize\ttfamily] at (0, 0) {initial-seeds};

    \node[font=\scriptsize, text=SteelBlue!85!black, anchor=south] at (0, 2.10)
      {$\mu_c$:~corpus enrichment};
    \node[font=\scriptsize, text=Orange!95!black, anchor=north] at (0, -2.10)
      {$\kappa_c$:~obligation reward};

    \node[font=\scriptsize\bfseries, text=black!85, anchor=west]
      at (3.05, 0.40) {Full Pipeline};

    \end{tikzpicture}%
    }
    \caption{Reachable behavior under a fixed exploration budget. An unguided fuzzer starting from \icode{riscv-tests}~\cite{riscvtests} reaches only the gray region; \tool's \textcolor{SteelBlue!85!black}{$\mu_c$} (top) and \textcolor{Orange!95!black}{$\kappa_c$} (bottom) expand reach, and \textcolor{cbpurple}{$\iota_c$} (ring) anchors witness injection to expose underconstraint bugs. Bug positions (\textcolor{red!75!black}{\boldmath$\times$}) are illustrative.}
    \label{fig:intro-insight}
\end{figure}

The triple plugs into a generic RL fuzzing loop at the three places it would otherwise default to randomness, choosing seeds, scoring an execution, and injecting into the witness (\autoref{fig:intro-insight}): $\mu_c$ constructs the initial set of test inputs (the corpus) (replacing random seed selection), $\kappa_c$ defines the reward that steers mutation choice, rewarding executions whose set of triggered obligations is new (replacing edge coverage), and $\iota_c$ anchors witness injection at the steps where an obligation is triggered (replacing infeasible blind witness injection). Thus $\mu_c$ and $\kappa_c$ address the first challenge by steering inputs toward corner cases, and $\iota_c$ addresses the second by making underconstraints observable.  These three hooks pair with two detectors mapped 1-to-1 onto the bug taxonomy: comparing final register values against a reference RISC-V emulator catches divergences, while $\iota_c$-anchored witness injection catches underconstraints invisible to output.  Our ablation (\autoref{sec:eval-rq2}) confirms each hook is necessary, with distinct roles: $\mu_c$ determines \emph{which region} of the zkVM behavior space the loop ever reaches (without it, the RL machinery has far fewer obligation-critical inputs to amplify); $\kappa_c$ \emph{redirects exploration} from new executor code paths to new obligation combinations, recovering corner cases that standard edge coverage misses; and $\iota_c$ is \emph{structurally required} for underconstraint bugs (the majority of our benchmark), which output comparison cannot see and blind witness injection rarely hits at production trace scale: with all three hooks removed, \tool detects only the 9 divergence bugs and none of the 28 underconstraints.

We implement this approach in \tool, evaluated on six leading production RV32IM zkVMs (spanning four distinct proof architectures).  On a curated benchmark of 37 independently sourced known bugs, \tool detects 25/37 (67.6\%) versus \arguzz's 5/37 (13.5\%), a $5.0\times$ improvement, with \tool's detections including all of \arguzz's.  \tool itself additionally discovers 14 previously unknown zero-day findings (all confirmed and patched).  Most detections are underconstraint bugs that require directly probing the constraint system, which obligation-anchored witness injection makes tractable within practical budgets.

\section{Background}\label{sec:background}

\subsection{Zero-Knowledge Proofs}\label{sec:bg-zk}

A \emph{zero-knowledge proof} allows one party (the \emph{prover}) to convince another party (the \emph{verifier}) that a statement is true, without revealing any information beyond the truth of the statement itself.  In the context of computation, the statement is: ``I executed this program on this input and obtained this output.''  The prover produces a short cryptographic \emph{proof} that the verifier can check in milliseconds, even if the original computation took hours.  If the proof system is \emph{sound}, no cheating prover can produce a valid proof for a false statement (e.g., a wrong output).  Concretely, proof systems require the statement to be expressed as an \emph{arithmetic circuit} (also called a \emph{proof circuit}), a fixed network of addition and multiplication gates over a finite field whose evaluation encodes the computation being proved.  Each application traditionally needed its own hand-crafted circuit, a labor-intensive and error-prone process that limited adoption to specialized use cases.

\subsection{Zero-Knowledge Virtual Machines}\label{sec:bg-zkvm}

A \emph{zero-knowledge virtual machine} (zkVM) applies zero-knowledge proofs to general-purpose program execution.  Rather than writing custom proof circuits for each application, a zkVM proves the correct execution of programs written in a standard instruction set, most commonly RISC-V, the open-source ISA used by all six systems we evaluate~\cite{risc0, sp1, jolt, openvm, nexus, pico}.

\autoref{fig:zkvm-pipeline} illustrates how a zkVM works, step by step:

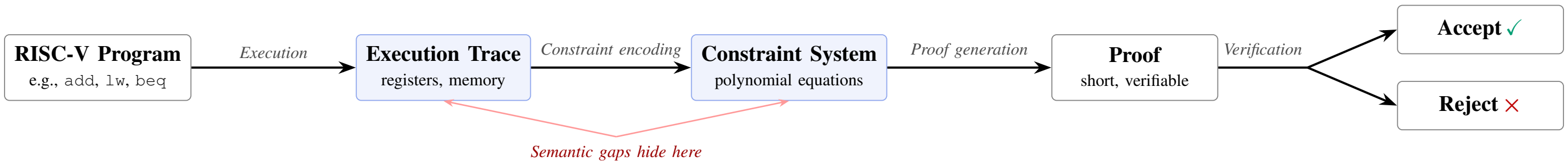
\begin{figure*}[t]
    \centering
    \resizebox{\linewidth}{!}{%
    \begin{tikzpicture}[
      A/.style={-{Stealth[length=2.5mm, width=1.8mm]}, line width=0.9pt},
      box/.style={draw=black!50, rounded corners=2pt, inner sep=4pt,
                  font=\small, align=center, minimum width=2.4cm, minimum height=0.95cm},
      ebox/.style={draw=RoyalBlue!60, fill=RoyalBlue!8, rounded corners=2pt, inner sep=4pt,
                  font=\small, align=center, minimum width=2.4cm, minimum height=0.95cm},
      sbox/.style={draw=black!50, rounded corners=2pt, inner sep=4pt,
                  font=\small, align=center, minimum width=2.4cm, minimum height=0.7cm},
      proc/.style={font=\scriptsize\itshape, text=black!70, align=center},
    ]
    \node[box]  (prog)   at (0,    0)     {\textbf{RISC-V Program}\\{\scriptsize e.g., \texttt{add}, \texttt{lw}, \texttt{beq}}};
    \node[ebox] (trace)  at (5.0,  0)     {\textbf{Execution Trace}\\{\scriptsize registers, memory}};
    \node[ebox] (constr) at (10.0, 0)     {\textbf{Constraint System}\\{\scriptsize polynomial equations}};
    \node[box]  (proof)  at (15.0, 0)     {\textbf{Proof}\\{\scriptsize short, verifiable}};
    \node[sbox] (accept) at (20.0, 0.55)  {\textbf{Accept}\,{\color{cbgreen}\checkmark}};
    \node[sbox] (reject) at (20.0, -0.55) {\textbf{Reject}\,{\color{red!75!black}$\boldsymbol{\times}$}};

    \draw[A] (prog)   -- node[proc, above] {Execution}           (trace);
    \draw[A] (trace)  -- node[proc, above] {Constraint encoding} (constr);
    \draw[A] (constr) -- node[proc, above] {Proof generation}    (proof);

    \coordinate (vfork) at (17.5, 0);
    \draw[line width=0.9pt] (proof.east) -- node[proc, above] {Verification} (vfork);
    \draw[A] (vfork) -- (accept.west);
    \draw[A] (vfork) -- (reject.west);

    \node[font=\scriptsize\itshape, text=red!60!black, anchor=north] (annot) at (7.5, -1.0) {Semantic gaps hide here};
    \draw[-{Stealth[length=1.5mm]}, red!40, line width=0.6pt] (annot.north) -- (trace.south);
    \draw[-{Stealth[length=1.5mm]}, red!40, line width=0.6pt] (annot.north) -- (constr.south);

    \end{tikzpicture}%
    }
    \caption{How a RISC-V zkVM works. Semantic gaps hide in the two highlighted artifacts: the execution trace and the constraint system.}
    \label{fig:zkvm-pipeline}
\end{figure*}

\begin{enumerate}[leftmargin=2em]
    \item \textbf{Execution.}  The zkVM's \emph{executor} runs the RISC-V program instruction by instruction, recording every intermediate state (register values, memory contents, and internal control signals) in a table called the \emph{execution trace}.  Each row of the table corresponds to one instruction step.

    \item \textbf{Constraint encoding.}  The execution trace, together with any auxiliary intermediate values the prover commits to, forms a \emph{witness} $w$.  The witness is encoded as a system of polynomial equations $A$ over a finite field $\mathbb{F}_p$ (the integers modulo a prime $p$); the constraint system \emph{accepts} iff every equation in $A$ evaluates to zero on $w$.  Each equation asserts a correctness property of one or more trace rows.  For example, for an \icode{add} instruction at row $i$, the constraint system must enforce:
    \[
    \mathit{trace}[i].\mathit{rd} = \mathit{trace}[i].\mathit{rs1} + \mathit{trace}[i].\mathit{rs2} \pmod{2^{32}}
    \]
    This encoding is called the \emph{arithmetization}.  Crucially, $\mathbb{F}_p$ does not natively model 32-bit hardware: overflow wrap-around, sign extension, range, and bitwise operations all have to be \emph{reconstructed} from field operations (typically by splitting each 32-bit value into byte-sized \emph{limbs} that are range-checked against a precomputed \emph{lookup} table of valid byte values), and every reconstruction is a place where the constraint system can silently deviate from the ISA.  All RISC-V zkVMs must enforce the same instruction semantics, regardless of how their internal trace tables are organized.

    \item \textbf{Proof generation.}  The prover uses a polynomial commitment scheme (e.g., FRI~\cite{fri} or KZG~\cite{kzg}) to produce a succinct proof that all constraints are satisfied simultaneously.

    \item \textbf{Verification.}  The verifier checks the proof without re-executing the program and without access to the trace or the program's internal state.
\end{enumerate}

\subsection{Where Bugs Hide: Constraint Obligations}\label{sec:bg-challenges}

A production zkVM's constraint system contains hundreds of thousands of polynomial equations. We call any deviation between the constraint system or executor and the ISA a \emph{semantic gap} (highlighted in \autoref{fig:zkvm-pipeline}); gaps cluster at non-trivial correctness properties we call \emph{constraint obligations}, rooted in two fixed sources: \textbf{(i)}~the ISA specification (what the instruction must compute, e.g., \icode{div} by zero, shift-amount masking) and \textbf{(ii)}~the proof system's algebraic structure (how it is encoded over $\mathbb{F}_p$, e.g., trace tables padded to power-of-two sizes, limb decomposition for range checks). \tool's methodology rests on this two-source derivation (\autoref{sec:trace-ir}). Two structural difficulties make these obligations hard to test:

\begin{itemize}
    \item \textbf{Rare trigger conditions.}  Obligations activate only under narrow conditions (division by zero, signed overflow, executions that overflow a power-of-two-sized table) almost never exercised by standard test suites or unguided fuzzing.

    \item \textbf{The oracle gap.}  A zkVM's own output cannot serve as ground truth, and the two failure modes call for different oracles. \emph{Divergence} bugs (executor off-spec on a legitimate input, a completeness gap) yield observable wrong register state and admit differential testing against a trusted RISC-V emulator. \emph{Underconstraint} bugs (executor correct but constraints too permissive, a soundness gap) leave the output identical and surface only as an alternative accepted witness; detecting them requires \emph{witness injection}, directly mutating the prover's intermediate values and re-checking acceptance.
\end{itemize}

\subsection{Reinforcement-Learning Fuzzing Baseline}\label{sec:bg-rl}

Modern fuzzers cast mutation-operator selection as a multi-armed bandit problem: at each iteration the fuzzer picks among a pool of operators (bit flip, byte splice, dictionary insertion, etc.) and updates that operator's expected reward from observed outcomes.  The UCB1 policy~\cite{ucb1} is the de facto default: it balances exploration and exploitation with a closed-form upper-confidence bound and has provable logarithmic regret in stationary settings.  \tool uses a UCB1 variant with a tunable exploration constant and an $\varepsilon$-greedy override (\autoref{sec:bandit-mutation}); we keep ``UCB1'' throughout when the distinction is immaterial.  Beyond the policy itself, a generic RL fuzzer ships with three default choices that operate where domain knowledge is absent: a random or developer-supplied seed corpus, edge coverage in the executor binary as the reward signal, and (for tools that probe internal state) blind perturbation of intermediate values.  All three defaults break on a zkVM (\autoref{sec:intro}); replacing each with a constraint-aware substitute is the central design problem \tool addresses (\autoref{sec:framework-overview}).

\subsection{Threat Model}\label{sec:bg-threat}

We adopt a \emph{malicious prover} model: the prover knows the zkVM source code, witness generation logic, and constraint system, and aims to produce a proof accepted by an honest verifier for an execution trace that deviates from correct RISC-V semantics.

\paragraph{Soundness}
Let $O$ denote a reference RISC-V oracle and $B$ a zkVM backend with executor $E_B$ and constraint system $A$ over $\mathbb{F}_p$.  Backend $B$ is \emph{sound} if, for all efficient provers and inputs $x$, any accepted proof encodes a valid execution of $x$ under $O$.  A \emph{soundness bug} is an input $x$ with a witness $w$ such that $A(w)$ accepts but $w$ encodes a computation different from $O(x)$.  We target bugs in the constraint system $A$ and the executor $E_B$, not in the underlying cryptographic proof backend.

\paragraph{Completeness}
Conversely, $B$ is \emph{complete} if every valid execution of $x$ admits at least one witness $w$ such that $A(w)$ accepts.  A \emph{completeness bug} is an input $x$ for which the honest witness (the one $E_B$ would produce in lockstep with $O$) is unavailable or rejected (e.g., $E_B(x)$ crashes or produces a register state inconsistent with $O(x)$).  Both soundness and completeness bugs are semantic gaps and are in scope for \tool.

\paragraph{Operational categorization}
The two formal classes map 1-to-1 to two operational categories: soundness bugs are \textbf{underconstraint} (detected via witness injection at the obligation step), completeness bugs are \textbf{divergence} (detected via differential register comparison against $O$). This mapping underpins the dual-detector design (\autoref{sec:framework-overview}) and the formal characterization in \autoref{sec:formal-properties}.

\paragraph{ISA scope}
\tool models RV32IM as specified in the RISC-V ISA manual~\cite{riscvspec}.  Misaligned accesses trap; division follows the specification exactly (division by zero yields quotient $-1$ and remainder equal to the dividend; signed overflow $-2^{31} \div -1$ yields quotient $-2^{31}$ and remainder $0$).  These corner cases are where several benchmark bugs manifest (\autoref{sec:eval-rq1}).

\section{Overview}\label{sec:overview}

We describe the framework architecture (\autoref{sec:framework-overview}) and walk through a concrete bug as a running example (\autoref{sec:motivating-example}); the formal characterization is in \autoref{sec:formal-properties}.
                                                                                       
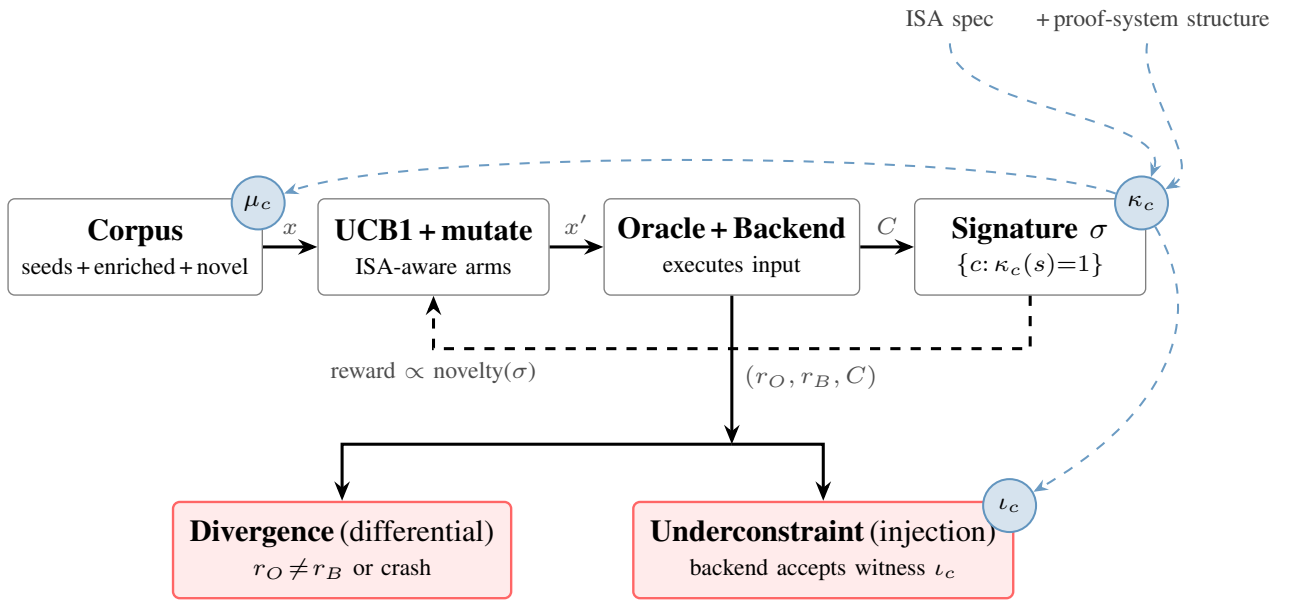
\begin{figure}[t]
    \centering
    \resizebox{0.96\columnwidth}{!}{%
    \begin{tikzpicture}[
      A/.style={-{Stealth[length=2mm, width=1.6mm]}, line width=0.9pt},
      DA/.style={-{Stealth[length=2mm, width=1.6mm]}, line width=0.9pt, dashed},
      hk/.style={-{Stealth[length=1.5mm, width=1.2mm]}, line width=0.6pt, draw=SteelBlue!80, dashed},
      hd/.style={draw=black!50, rounded corners=2pt, inner sep=4pt,
                 font=\small, align=center, minimum width=2.40cm, minimum height=1.00cm},
      dbox/.style={draw=red!60, fill=red!8, rounded corners=2pt, inner sep=5pt, line width=0.7pt,
                   font=\small, align=center, minimum width=3.10cm, minimum height=1.00cm},
      hbadge/.style={circle, draw=SteelBlue!85, fill=SteelBlue!22, line width=0.6pt,
                     font=\scriptsize, inner sep=0.5pt, minimum size=5.5mm},
      lbl/.style={font=\scriptsize, text=black!65},
      iolbl/.style={font=\scriptsize, text=black!70},
    ]
    \node[font=\scriptsize, text=black!75, anchor=east] (isa)
      at ( 4.40, 2.55) {ISA spec};
    \node[font=\scriptsize, text=black!75, anchor=west] (psys)
      at ( 4.55, 2.55) {\,+\,proof-system structure};

    \node[hd]  (corpus) at (-4.65, 0.20) {\textbf{Corpus}\\{\scriptsize seeds\,+\,enriched\,+\,novel}};
    \node[hd]  (bandit) at (-1.55, 0.20) {\textbf{UCB1\,+\,mutate}\\{\scriptsize ISA-aware arms}};
    \node[hd]  (oracle) at ( 1.55, 0.20) {\textbf{Oracle\,+\,Backend}\\{\scriptsize executes input}};
    \node[hd]  (sig)    at ( 4.65, 0.20) {\textbf{Signature $\sigma$}\\{\scriptsize $\{c{:}\,\kappa_c(s){=}1\}$}};

    \draw[A] (corpus) -- node[iolbl, above] {$x$} (bandit);
    \draw[A] (bandit) -- node[iolbl, above] {$x'$} (oracle);
    \draw[A] (oracle) -- node[iolbl, above] {$C$} (sig);
    \draw[DA] (sig.south) -- ++(0,-0.55)
              -| node[iolbl, below, pos=0.50] {reward $\propto$ novelty($\sigma$)} (bandit.south);

    \node[dbox] (det1) at (-2.50, -2.95)
      {\textbf{Divergence}\,(differential)\\{\scriptsize $r_O\!\neq\!r_B$ or crash}};
    \node[dbox] (det2) at ( 2.50, -2.95)
      {\textbf{Underconstraint}\,(injection)\\{\scriptsize backend accepts witness $\iota_c$}};

    \coordinate (jct) at (1.55, -1.85);
    \draw[A] (oracle.south) -- node[iolbl, right, pos=0.55] {$(r_O,r_B,C)$} (jct);
    \draw[A] (jct) -| (det1.north);
    \draw[A] (jct) -| (det2.north);

    \node[hbadge] (mu)  at ($(corpus.north east)+(-0.05,-0.05)$) {$\mu_c$};
    \node[hbadge] (ka)  at ($(sig.north east)   +(-0.05,-0.05)$) {$\kappa_c$};
    \node[hbadge] (iot) at ($(det2.north east)  +(-0.05,-0.05)$) {$\iota_c$};

    \draw[hk] (isa)  to[out=270, in=70, looseness=0.7] (ka);
    \draw[hk] (psys) to[out=260, in=30, looseness=0.8] (ka);
    \draw[hk] (ka)   to[out=160, in=20, looseness=0.40] (mu);
    \draw[hk] (ka)   to[out=300, in=30, looseness=1.10] (iot);

    \path (-6.10, -3.50) (6.10, 2.85);
    \end{tikzpicture}%
    }
    \caption{Architecture of \tool. Each obligation's predicate $\kappa_c$ is derived from the ISA spec and the proof-system structure, and $\mu_c$ and $\iota_c$ are derived from $\kappa_c$ (dashed arcs). The three plug into the fuzzing loop at the corpus, the reward signal, and the underconstraint detector.}
    \label{fig:overview}
\end{figure}
                                                                                                                                                                                                                   
\subsection{Framework Overview}\label{sec:framework-overview}

\autoref{fig:overview} illustrates \tool's architecture: an RL fuzzing loop in which the obligation triple $(\mu_c, \kappa_c, \iota_c)$ replaces three default choices in a generic fuzzer. Each of the 60 obligations enumerated by the semantic model (\autoref{sec:trace-ir}) supplies an enrichment template $\mu_c$ (inputs reaching the trigger condition), a predicate $\kappa_c$ (recognizes activation on a trace), and an injection template $\iota_c$ (injects an alternative witness at the activation point).

\paragraph{The three obligation hooks}
$\mu_c$ enriches each \icode{riscv-tests} seed into an input that exercises one obligation from the first mutation iteration onward (\autoref{sec:initial-corpus}). $\kappa_c$ projects each post-execution trace into an obligation signature whose novelty rewards the bandit arm that selected the mutation, replacing edge coverage (uncorrelated with constraint structure) with constraint-state novelty (\autoref{sec:signature-feedback}). $\iota_c$ anchors witness injection at the activation step $(c,s)$ and probes whether the constraint system accepts an alternative encoding (\autoref{sec:witness-search}); without anchoring, the search is infeasible.

\paragraph{Bandit-guided ISA-aware mutation}
At each iteration, a UCB1 multi-armed bandit selects an ISA-aware mutation arm (e.g., register swap, boundary-immediate substitution; full set in \autoref{tbl:mutation-arms}) and applies it to a corpus entry (\autoref{sec:bandit-mutation}). Several arms incorporate obligation-specific heuristics, while obligation-signature novelty provides feedback for subsequent arm selection.

\paragraph{Oracle, backend, and two detectors}
Each mutated input runs in parallel on a reference RV32IM simulator (yielding $r_O$; \autoref{sec:oracle}) and on the instrumented backend (yielding $r_B$ and the obligation activations $C$). Two detectors share the loop and target the two semantic-gap classes (\autoref{sec:bg-threat}): differential register comparison flags \emph{divergence} bugs ($r_O \neq r_B$ or executor crash; crashes routed through assertion-bypass instrumentation, \autoref{sec:instrumentation}); $\iota_c$-anchored witness injection at $(c,s)\in C$ flags \emph{underconstraint} bugs. The two detectors are orthogonal: divergence bugs leave no alternative witness for injection to find, and underconstraint bugs leave no register mismatch for differential to flag.
                                                                                                                                                                                                                   
\subsection{Motivating Example}\label{sec:motivating-example}                          
                                                                                                                                                                                                                   
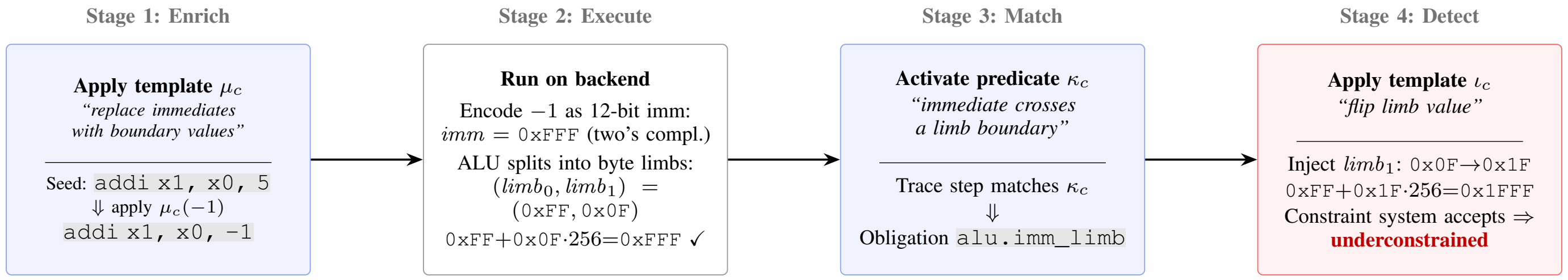
\begin{figure*}[t]
\centering
\resizebox{\textwidth}{!}{%
\begin{tikzpicture}[
  A/.style={-{Stealth[length=2.2mm, width=1.7mm]}, line width=0.8pt},
  box/.style={draw=black!40, rounded corners=2pt, inner sep=4pt,
              font=\footnotesize, align=center, text width=3.8cm,
              minimum height=3.1cm, anchor=north},
  sbox/.style={draw=RoyalBlue!60, fill=RoyalBlue!8, rounded corners=2pt,
               inner sep=4pt, font=\footnotesize, align=center, text width=3.8cm,
               minimum height=3.1cm, anchor=north},
  dbox/.style={draw=red!60, fill=red!5, rounded corners=2pt,
               inner sep=4pt, font=\footnotesize, align=center, text width=3.8cm,
               minimum height=3.1cm, anchor=north},
  phase/.style={font=\footnotesize\bfseries, text=black!55, anchor=south},
]

\node[phase] at (0,    1.45) {Stage 1: Enrich};
\node[phase] at (5.6,  1.45) {Stage 2: Execute};
\node[phase] at (11.2, 1.45) {Stage 3: Match};
\node[phase] at (16.8, 1.45) {Stage 4: Detect};

\node[sbox] (p1) at (0, 1.35) {%
  \textbf{Apply template $\mu_c$}\\[1pt]
  {\scriptsize
  \emph{``replace immediates}\\
  \emph{with boundary values''}\\[2pt]
  \rule{0.8\linewidth}{0.3pt}\\[2pt]
  Seed: \icode{addi x1, x0, 5}\\[1pt]
  $\Downarrow$~apply $\mu_c(-1)$\\[1pt]
  \icode{addi x1, x0, -1}}%
};

\node[box] (p2) at (5.6, 1.35) {%
  \textbf{Run on backend}\\[3pt]
  Encode $-1$ as 12-bit imm:\\
  $\mathit{imm} = \texttt{0xFFF}$ (two's compl.)\\[2pt]
  ALU splits into byte limbs:\\
  $(\mathit{limb}_0,\mathit{limb}_1) = (\texttt{0xFF},\texttt{0x0F})$\\[2pt]
  $\texttt{0xFF}{+}\texttt{0x0F}{\cdot}256{=}\texttt{0xFFF}$~\checkmark
};

\node[sbox] (p3) at (11.2, 1.35) {%
  \textbf{Activate predicate $\kappa_c$}\\[1pt]
  \emph{``immediate crosses}\\
  \emph{a limb boundary''}\\[2pt]
  \rule{0.8\linewidth}{0.3pt}\\[2pt]
  Trace step matches $\kappa_c$\\[1pt]
  $\Downarrow$\\[1pt]
  Obligation \icode{alu.imm\_limb}
};

\node[dbox] (p4) at (16.8, 1.35) {%
  \textbf{Apply template $\iota_c$}\\[1pt]
  \emph{``flip limb value''}\\[2pt]
  \rule{0.8\linewidth}{0.3pt}\\[2pt]
  Inject $\mathit{limb}_1$: \texttt{0x0F}$\to$\texttt{0x1F}\\[1pt]
  $\texttt{0xFF}{+}\texttt{0x1F}{\cdot}256{=}\texttt{0x1FFF}$\\[1pt]
  Constraint system accepts $\Rightarrow$\\
  \textcolor{red!75!black}{\textbf{underconstrained}}
};

\draw[A] (p1.east) -- (p2.west);
\draw[A] (p2.east) -- (p3.west);
\draw[A] (p3.east) -- (p4.west);

\end{tikzpicture}%
}
\caption{\tool finds a missing range check in OpenVM's ALU chip via obligation \icode{alu.imm\_limb}: $\mu_c$ places immediate $-1$ in the initial corpus, $\kappa_c$ fires at the ALU step, and the constraints accept $\iota_c$'s alternative limb decomposition (\texttt{0x1FFF} $\neq$ \texttt{0xFFF}).}
\label{fig:motivating}
\end{figure*}                                                                                                                                                                                                      
                                                                                                                                                                                                                   
We illustrate the three obligation-triple components with a real constraint bug in OpenVM's ALU chip (\autoref{fig:motivating}), drawn from our benchmark of 37 known bugs in \autoref{sec:eval-rq1}: a missing range check that lets a malicious prover claim an alternative encoding of a 12-bit immediate value that the ISA never permits.

\paragraph{Background: why immediates are split into bytes}
As described in \autoref{sec:bg-zkvm}, zkVMs split values into byte-sized limbs that are range-checked against a lookup table; the 12-bit RISC-V immediate of \icode{addi} is then \emph{recomposed} from these limbs.

In OpenVM, the ALU chip splits each 12-bit immediate into two byte limbs $(\mathit{limb}_0, \mathit{limb}_1)$ and enforces a \emph{recomposition constraint}:
\begin{equation}\label{eq:recomp}
\mathit{limb}_0 + \mathit{limb}_1 \cdot 256 = \mathit{immediate}
\end{equation}
For instance, the immediate $-1$ (encoded as \icode{0xFFF} in 12-bit two's complement) decomposes as $(\mathit{limb}_0, \mathit{limb}_1) = (\texttt{0xFF}, \texttt{0x0F})$, satisfying \autoref{eq:recomp}.  Because \autoref{eq:recomp} alone does not constrain how the prover chooses the limbs, the decomposition is unique only if the upper limb is range-checked to its actual 4-bit width via a range-check lookup:
\begin{equation}\label{eq:range}
0 \leq \mathit{limb}_0 < 256, \quad 0 \leq \mathit{limb}_1 < 16
\end{equation}                                                                                                                                                                                                     
                                          
\paragraph{The bug}                                                                                                                                                                                                
Consider the instruction \icode{addi x1, x0, -1}, which sets register \icode{x1} to $-1$ (i.e., \icode{0xFFFF\_FFFF} in unsigned 32-bit representation). OpenVM's ALU chip uses the recomposition (\autoref{eq:recomp}) as the effective immediate without separately checking it against the decoded 12-bit value, and \emph{omitted} the upper-limb range-check constraint (\autoref{eq:range}) that bounds $\mathit{limb}_1 < 16$. With $\mathit{limb}_1$ unconstrained beyond byte width, a malicious prover can supply $(\texttt{0xFF}, \texttt{0x1F})$, which the constraint system accepts and recomposes to $\texttt{0xFF} + \texttt{0x1F} \cdot 256 = \texttt{0x1FFF}$. The ALU chip then uses \icode{0x1FFF} in place of \icode{0xFFF} when computing the instruction result, forging a proof for an execution that never happened.                                                          
                                                                                                                                                                                                                   
\paragraph{How \tool discovers this bug}
The \icode{alu.imm\_limb} obligation specifies a triple: enrichment template $\mu_c$ (``replace immediates with boundary values''), predicate $\kappa_c$ (``immediate crosses a limb boundary''), and injection template $\iota_c$ (``flip limb value''). \tool applies $\mu_c$ to a seed containing \icode{addi}, producing \icode{addi x1, x0, -1} as part of the initial corpus. When the input runs, the backend trace activates predicate $\kappa_c$ at the ALU step. At that step, $\iota_c$ rewrites the limb pair to an alternative decomposition ($\mathit{limb}_1$: \texttt{0x0F} $\to$ \texttt{0x1F}) and re-evaluates the constraint system. The constraints accept the alternative, confirming an underconstrained witness. Because $\mu_c$ places the trigger input in the initial corpus, no combinatorial mutation is required; the bug is detected once the loop selects this input.

\paragraph{Where the RL loop matters}
Not all bugs yield to a single $\mu_c$.  When a bug requires the simultaneous activation of multiple obligations or a specific instruction-sequence pattern that no single enrichment template produces, $\kappa_c$ and $\iota_c$ carry the load: as the bandit explores combinatorial mutations of the enriched corpus, $\kappa_c$ rewards inputs that activate previously unseen obligation combinations, and once the right pattern fires, $\iota_c$ injects an alternative witness at the activation site and detects the underconstraint.  Our ablation (\autoref{sec:eval-rq2}) quantifies each component's contribution.

\section{Constraint Obligations}\label{sec:trace-ir}

RV32IM zkVMs must enforce the correctness properties required by the RISC-V semantics. Register x0 must remain zero. Immediate values must decompose into correctly range-checked byte-sized pieces. Shift amounts must be masked to 5~bits. Memory timestamps must increase monotonically. We call each such property a \emph{constraint obligation}: a correctness condition that every faithful zkVM must enforce. When a backend fails to enforce an obligation, a semantic gap exists: an underconstraint bug if the constraint system admits an alternative witness, or a divergence bug if the executor itself deviates from the ISA on a legitimate input.

This insight is the foundation of \tool's semantic model. Rather than analyzing each backend's constraint system directly, we systematically enumerate the constraint obligations that any correct RV32IM zkVM must satisfy. Each obligation $c$ is defined by a \emph{predicate} $\kappa_c$ on a trace step that returns whether the obligation is activated; $\kappa_c$ is the obligation's identity. The two other components of the triple, the \emph{enrichment template} $\mu_c$ and the \emph{injection template} $\iota_c$, are mechanically derived from $\kappa_c$: $\mu_c$ inverts the predicate's trigger condition to construct inputs that satisfy it, and $\iota_c$ targets the witness cells named in the predicate to probe the constraint system. Both derived components are consumed by the fuzzing loop in \autoref{sec:fuzzing-framework}; this section focuses on the obligations themselves, their derivation, predicate form, and coverage.

\subsection{Obligation Derivation}\label{sec:semantic-model}

We derive the 60 constraint obligations through a two-step top-down methodology: \emph{functional decomposition} identifies the constraint components, and \emph{obligation enumeration} identifies the correctness properties within each component.

\paragraph{Step 1: Functional decomposition}
We organize the obligation set around the functional components commonly found in production RV32IM zkVMs: an instruction \emph{decoder} (field extraction from 32-bit words), an \emph{ALU} (arithmetic and logic), an \emph{arithmetic unit} (multiplication and division), \emph{control flow} (branches, jumps, system calls), an \emph{execution pipeline} (operand routing and result binding), a \emph{memory subsystem} (load/store consistency), \emph{lookup tables} (range checks, boolean and XOR multiplicities), cross-component \emph{interactions} (digest routing), \emph{row management} (padding and table lifecycle), and a \emph{timestamp model} (ordering guarantees). All 10 components are present in each of the six target backends (\autoref{tbl:backends}), giving 10 functional groups (\autoref{tbl:obligations}); a backend that omitted one of them would effectively leave an ISA function unconstrained, which is itself an underconstraint.

\paragraph{Step 2: Obligation enumeration}
Within each functional group, we enumerate obligations along the two-source taxonomy of \autoref{sec:bg-challenges}:

\begin{itemize}
    \item \textbf{Semantic} (ISA-rooted): x0~immutability (Decode), shift-amount masking to 5~bits (ALU), the division-by-zero result convention (Arithmetic), signed-vs-unsigned branch selection (Control), and similar properties whose violation makes the arithmetization accept a trace that encodes an incorrect RV32IM execution.

    \item \textbf{Algebraic} (STARK-rooted): every backend pads trace tables, segment tables, memory regions, lookup accumulators, commitment vectors, and range-check tables to powers of two (required by the FFT/NTT evaluation domain in STARK-based systems; analogous power-of-two structural constraints arise in Lasso's subtable decomposition and Nova's step-circuit padding), so crossing a $2^k$ threshold forces constraint-table reallocation that exercises logic dormant at smaller sizes. These obligations live in the functional groups they affect (e.g., the memory table's $2^k$ capacity threshold is a Memory obligation, not a separate ``resource'' concern).
\end{itemize}

\noindent Both sources are enumerated uniformly. \autoref{tbl:obligations} lists all 10 groups, their counts, and one representative obligation with its predicate per group; the full enumeration of all 60 obligations and their linked benchmark bugs appears in Appendix~\ref{app:obligations}.

\paragraph{Non-triviality filter}
At every scope of enumeration we apply a non-triviality filter: a candidate is admitted only when its underlying source (an ISA semantic edge case or a STARK-structural component) introduces non-trivial constraint complexity in the $\mathbb{F}_p$ embedding. Trivial cases are excluded by construction. The filter applies pervasively, at three scales:
\begin{itemize}
    \item \emph{Instruction level}: RV32IM features uniformly trivial across backends are not enumerated. For example, EBREAK and FENCE are implemented as halt/no-op across all six target backends with no constraint-system logic, and \tool does not introduce obligations for them.
    \item \emph{Operand level}: operand configurations that exercise no constraint complexity are not retained as test inputs. For example, \icode{add x1, x2, 0} or mid-range non-boundary immediates exercise no per-limb decomposition stress and contribute no test value to any ALU obligation.
    \item \emph{Trace-state level}: traces that do not approach algebraic boundaries are filtered before the obligations targeting those boundaries. For example, short programs that do not cross any $2^k$ threshold do not exercise table-padding obligations.
\end{itemize}
\noindent The filter is what makes systematic enumeration feasible: the raw configuration space is exponential, but its non-trivial subset is bounded by the structure of the ISA and STARK arithmetization combined.

\paragraph{Scope}
Beyond the trivial cases removed by the filter above, two further classes are out of scope: implementation-specific properties such as backend-chosen memory range bounds, which are not ISA-level invariants and so are not enumerated; and the cryptographic proof backend (FRI / IPA / Lasso / Nova), which is assumed sound consistent with the threat model in \autoref{sec:bg-threat}.

\paragraph{Running example}
Consider \icode{addi x1, x0, -1} executed on a production zkVM. The ALU component decomposes the 12-bit immediate ($-1$, encoded as \icode{0xFFF}) into byte-sized pieces for constraint evaluation over a finite field. At this trace step, \tool evaluates the \icode{alu.imm\_limb} predicate, which checks whether the immediate value crosses a byte boundary. The same predicate applies identically on any other backend processing the same instruction: obligations are defined on backend-agnostic trace-step properties, abstracting away how each backend internally represents the decomposition.

\paragraph{Formal definition}
A trace step $s \in \mathit{Steps}(x)$ records the step index, instruction fields, and computed values (register state, immediate, memory address, etc.) during execution of input $x$ on a backend. Each constraint obligation $c \in \mathcal{C}$ is identified by a predicate $\kappa_c : \mathit{Steps} \to \{0, 1\}$ that returns 1 when the trace step exercises the obligation. The \emph{obligation signature} for input $x$ is the set of all activated obligations:
\begin{equation}\label{eq:signature}
\sigma(x) = \bigl\{c \in \mathcal{C} \;\big|\; \exists\, s \in \mathit{Steps}(x) : \kappa_c(s) = 1\bigr\}
\end{equation}
The signature is what the fuzzer observes about an input; \autoref{sec:fuzzing-framework} describes how it, together with each activated obligation's enrichment and injection templates, is consumed by the fuzzing loop.

\paragraph{Predicate form}
Each predicate $\kappa_c$ is a pattern-matching rule on the trace step's instruction class and computed values, combining an instruction-class filter with a condition on the step's computed values (e.g., a value crossing a $2^k$ threshold, a register index at a special position, or a field element near the modulus). \autoref{tbl:obligations} gives one representative per group; the full set of 60 predicates follows this pattern (Appendix~\ref{app:obligations}).

\paragraph{Provenance and validation}
We validate the set against 37 known bugs drawn from professional security audits, prior dynamic testing~\cite{arguzz}, and public issue trackers (with overlap; see \autoref{sec:eval-rq1}). Of the 60 obligations, 22 are linked to at least one of the 25 detected benchmark bugs (Appendix~\ref{app:obligations}); the remaining obligations have no linked detection in this evaluation. The evaluation additionally includes 14 zero-day findings discovered by \tool and held out from obligation derivation. All 12 missed known bugs require cross-component reasoning beyond single-step obligations: five require coordinated modifications to multiple witness fields, and seven involve backend-specific structures (e.g., SP1's recursion VM) that the common obligation set does not cover.

\paragraph{Coverage and extensibility}
The 60 obligations cover the components present across our six target backends and, within each component, the correctness properties identified through ISA and arithmetization analysis: semantic obligations target ISA edge cases (bit-width truncation, special-case results, signed/unsigned disambiguation), and algebraic obligations target $2^n$-threshold transitions at the tables a STARK-based backend pads. Each obligation is a single-step predicate, so bugs that depend on cross-component reasoning fall outside this framing (\autoref{sec:eval-rq1}). New obligations, including backend-specific ones, enter the framework as additional trace-step predicates $\kappa_c$, with no changes to the fuzzing loop, bandit, or witness injection. Deriving the 60 obligations took one researcher about one week; encoding them and instrumenting each backend took about ten hours and 200--600 lines of Rust per backend. Porting \tool to another ISA requires re-deriving the ISA-specific obligations and adapting the instruction mutator and the reference oracle; the RL loop, reward, and injection machinery carry over unchanged.

\begin{table}[t]
\centering
\small
\begin{tabular}{lll}
\toprule
\textbf{Backend} & \textbf{Proving System} & \textbf{Architecture} \\
\midrule
SP1     & STARK (Plonky3) & Modular chip AIR \\
Pico    & STARK (Plonky3) & Monolithic STARK machine \\
Risc0   & STARK (FRI)     & Circuit-rv32im \\
Jolt    & Lookup (Lasso)  & Lookup-based verification \\
Nexus   & Nova (IVC)      & Incremental verification \\
OpenVM  & STARK (Plonky3) & Modular chip AIR \\
\bottomrule
\end{tabular}
\caption{Supported zkVM backends.}
\label{tbl:backends}
\end{table}

\begin{table*}[t]
\centering
\small
\setlength{\tabcolsep}{4pt}
\begin{tabular}{llp{3.9cm}p{7.4cm}}
\toprule
\textbf{Group (\#)} & \textbf{Example Obligation} & \textbf{Description} & \textbf{Predicate $\kappa_c(s)$} \\
\midrule
Decode (7)      & \icode{decode.zero\_reg}        & x0 immutability                 & $s.\mathit{rd} = \texttt{x0} \land s$~is a write \\
ALU (8)         & \icode{alu.shift\_mask}         & shift-amount masking            & $s$~is shift $\land\, s.\mathit{shamt} \geq 32$ \\
Arithmetic (6)  & \icode{arith.div\_rem}          & division remainder bounds       & $s$~is div/rem $\land\, (s.\mathit{divisor}{=}0 \lor s.\mathit{dividend}{=}{-}2^{31})$ \\
Control (8)     & \icode{control.branch\_cond}    & branch condition selection      & $s$~is branch $\land\, s.\mathit{cmp\_signed} \neq s.\mathit{cmp\_unsigned}$ \\
Execution (7)   & \icode{exec.operand\_bind}      & source/dest operand binding     & $s.\mathit{rs1} = s.\mathit{rs2}$ \\
Memory (12)     & \icode{mem.addr\_align}         & address alignment               & $s$~is memory $\land\, s.\mathit{addr} \bmod 4 \neq 0$ \\
Lookup (5)      & \icode{lookup.boolean\_mult}    & boolean multiplicity            & $s$~is lookup interaction step \\
Interaction (2) & \icode{interaction.digest}      & cross-component digest routing  & $s$~is interaction $\land\, s.\mathit{kind}_1 \neq s.\mathit{kind}_2$ \\
Row (2)         & \icode{row.padding\_leak}       & padding row interaction leakage & $s$~is padding row \\
Time (3)        & \icode{time.boundary\_origin}   & timestamp boundary origin       & $s$~is first timestamp step (init row) \\
\bottomrule
\end{tabular}
\caption{Constraint obligations by functional group, with one representative predicate $\kappa_c$ per group. All 60 obligations are listed in Appendix~\ref{app:obligations}.}
\label{tbl:obligations}
\end{table*}

\subsection{How the Model Catches Bugs}\label{sec:category-examples}

\autoref{tbl:obligations} illustrates the model's operation. Beyond the motivating example (\autoref{sec:motivating-example}), the algebraic obligations within each group are particularly effective for divergence bugs: the Memory group's capacity threshold catches a memory-size crash at a $2^k$ address boundary, and the Execution group's segment boundary catches an executor failure when the trace crosses a $2^k$ segment boundary. In each case, the obligation names the condition and the detector confirms the bug within the same loop iteration. Further zero-day case studies are in \autoref{sec:eval-rq3}.

\subsection{Formal Properties of the Semantic Model}\label{sec:formal-properties}

We formalize the properties that underpin the semantic model's detection guarantees and the complementary roles of the two detectors.

\begin{definition}[Semantic Gap]\label{def:semantic-gap}
Let $B$ be a zkVM backend with executor $E_B$ and constraint system $A$ over field $\mathbb{F}_p$, let $O$ be the reference RISC-V oracle, and let $x$ be an input program. Write $r_t \in \mathbb{Z}_{2^{32}}^{32}$ for the register state after step $t$ and $w_t \in \mathbb{F}_p^m$ for the witness row at step $t$ (where $m$ is the trace width). A \emph{semantic gap} exists at instruction step $t$ of $x$ in $B$ if at least one of the following holds:
\begin{itemize}
    \item \textbf{Divergence:} The executor computes a register state that differs from the oracle: $E_B(x, t) \neq O(x, t)$, i.e., $\exists\, j \in [32] : r^B_{t,j} \neq r^O_{t,j}$.  In other words, $E_B$ deviates from the ISA on a legitimate $x$, so no spec-conforming witness exists for $A$ to accept.
    \item \textbf{Underconstraint:} The executor computes the correct state ($E_B(x, t) = O(x, t)$), but the constraint system admits an alternative witness $w' \neq w_t$ at step $t$ such that $A(w') = \mathsf{accept}$ and $\mathit{decode}(w') \neq \mathit{decode}(w_t)$, where $\mathit{decode}: \mathbb{F}_p^m \to \mathbb{Z}_{2^{32}}^{32}$ extracts the register state encoded by a witness row.  In other words, $A$ accepts an invalid witness, and a malicious prover could substitute $w'$ while still passing verification.
\end{itemize}
\end{definition}

\noindent Divergence gaps arise from prover-side (executor) errors; underconstraint gaps arise from verifier-side (constraint system) errors. The two classes are operationally disjoint by definition and map 1-to-1 to the two formal failure modes (completeness, soundness) of \autoref{sec:bg-threat}. The $\mathit{decode}$ function is backend-specific (e.g., one backend extracts 32~field elements and reduces modulo $2^{32}$; another reconstructs from lookup table indices), but the definition is parametric in this mapping. The two classes require different detection mechanisms: divergence via output comparison, underconstraint via alternative-witness construction.

\paragraph{Dual-detector coverage characterization}
Both detectors run on every evaluation. Their coverage properties with respect to Definition~\ref{def:semantic-gap} are as follows:

\begin{enumerate}
    \item \textbf{The differential detector covers divergence gaps.} Register comparison detects all divergence gaps whose deviation propagates to the final register state, by construction (comparison is exact). Divergence gaps that crash the executor are handled via assertion-bypass instrumentation (\autoref{sec:instrumentation}), which converts fatal assertions to logged warnings and allows execution to continue to a (corrupted) final state.

    \item \textbf{The injection detector covers underconstraint gaps within instrumentation reach.} Witness injection can detect underconstraint gaps at any step $t$ for which an obligation activation exists at step $s$ such that $-\delta_- \le t - s \le \delta_+$ (the asymmetric injection search window, default $(\delta_-, \delta_+) = (16, 64)$; the asymmetry reflects that algebraic effects of an alternative witness propagate forward through limb chains, lookup multiplicities, and timestamp ordering, leaving more candidate cells downstream of the activation than upstream). The coverage boundary is twofold: the window limits spatial reach, and the obligation model limits semantic reach (gaps at uninstrumented constraint components are invisible to injection).

    \item \textbf{Neither detector subsumes the other.} Divergence gaps (executor errors) produce no alternative witness and thus cannot be detected by injection. Underconstraint gaps (correct executor, permissive constraints) produce no register mismatch and thus cannot be detected by differential comparison. This separation is structural: on the known benchmark, of the 25 bugs detected by \tool, 9 are divergence-side and 16 are underconstraint-side, confirming that the two detectors are orthogonal capabilities targeting orthogonal classes.\label{item:separation}
\end{enumerate}

\paragraph{Detector complementarity bound}\label{cor:complementarity}
On the 37 known bugs, removing the injection detector reduces \tool's detections from 25/37 (67.6\%) to at most 9/37 (the divergence bugs). This bound is structural (it holds regardless of time budget, seed corpus, or mutation strategy) because underconstraint gaps produce no observable signal in differential comparison. This observation quantifies the precise contribution of witness injection and explains the performance gap between \tool and prior dynamic testers such as \arguzz. \arguzz's fault injection targets the same underconstraint gaps; absent an obligation model to anchor the injection site, reaching equivalent coverage requires substantially more time and compute.

\subsection{Backend Instrumentation}\label{sec:instrumentation}

\tool instruments each backend's trace-generation path via automated source patching in three passes: (1)~\textbf{infrastructure injection} adds a uniform obligation-evaluation API; (2)~\textbf{assertion bypass} replaces Rust \icode{panic!}/\icode{assert!} macros (identified by textual pattern matching on source code) with logged warnings, enabling execution of crash-inducing inputs (sandboxed execution with a watchdog mitigates the UB risk from bypassed assertions); (3)~\textbf{predicate hooks} are inserted at each backend's constraint-component sites (ALU chips evaluate limb-decomposition predicates, memory chips evaluate address-alignment predicates, etc.).

Obligation evaluations faithfully reflect computed trace facts, but coverage is limited to instrumented constraint components: a new component requires a new predicate hook. Each backend exposes a uniform interface returning 32 registers and activated obligations. \autoref{tbl:backends} lists the six supported backends.

\section{Obligation-Targeted Semantic Fuzzing}\label{sec:fuzzing-framework}

Building on the semantic model of \autoref{sec:trace-ir}, this section details how the obligation triple plugs into the loop sketched in \autoref{sec:framework-overview}: $\mu_c$ supplies the initial corpus (\autoref{sec:initial-corpus}), $\kappa_c$ defines the bandit reward (\autoref{sec:signature-feedback}), and $\iota_c$ anchors witness injection (\autoref{sec:witness-search}).

\subsection{The Fuzzing Loop}\label{sec:fuzzing-loop}

\autoref{alg:fuzzing} presents the loop as five procedures whose definitions are deferred to subsequent subsections. After \textsc{InitialCorpus} populates the corpus from an obligation-enriched seed set, each iteration mutates an input (\textsc{Mutate}), executes it on the oracle and the instrumented backend (\textsc{OracleRun}, \textsc{BackendRun}, returning $r_O, r_B$, and the activation list $C$), grows the corpus on novel obligation signatures, updates the bandit, and probes the witness at obligation-anchored steps (\textsc{WitnessProbe}). Two detectors share the loop on the same $C$: the \textbf{differential detector} (\autoref{sec:differential-detector}) compares registers to flag divergences; the \textbf{injection detector} (\autoref{sec:witness-search}) runs \textsc{WitnessProbe} to flag underconstraints.

\begin{algorithm}[t]
    \caption{The \tool Fuzzing Loop}
    \begin{algorithmic}[1]
        \Procedure{\textsc{Fuzz}}{seeds $S$, backend $B$, oracle $O$, iterations $N$}
            \State $\mathit{corpus} \leftarrow \textsc{InitialCorpus}(S)$
            \State $\mathit{Seen} \leftarrow \emptyset$;\ \ $\mathit{bugs} \leftarrow \emptyset$;\ \ $\mathcal{B} \leftarrow \textsc{UCB1}(\{0,\dots,7\})$
            \For{$i = 1$ \textbf{to} $N$}
                \State $(x', a) \leftarrow \textsc{Mutate}(\mathit{corpus}, \mathcal{B})$
                \State $r_O \leftarrow \textsc{OracleRun}(x', O)$
                \State $(r_B, C) \leftarrow \textsc{BackendRun}(x', B)$
                \If{$r_O \neq r_B$} \Comment{\textbf{differential detector}: divergence}
                    \State $\mathit{bugs} \leftarrow \mathit{bugs} \cup \{(x', \texttt{"divergence"})\}$
                \EndIf
                \State $\sigma \leftarrow \textsc{Signature}(C)$
                \State $r \leftarrow \textsc{Reward}(\sigma, \mathit{Seen})$
                \If{$\sigma \notin \mathit{Seen}$}
                    \State $\mathit{corpus} \leftarrow \mathit{corpus} \cup \{x'\}$;\ \ $\mathit{Seen} \leftarrow \mathit{Seen} \cup \{\sigma\}$
                \EndIf
                \State $\mathcal{B}.\textsc{update}(a, r)$
                \State $\mathit{bugs} \leftarrow \mathit{bugs} \cup \textsc{WitnessProbe}(C, B)$ \Comment{\textbf{injection detector}: underconstraint}
            \EndFor
            \State \Return $\mathit{bugs}$
        \EndProcedure
    \end{algorithmic}
    \label{alg:fuzzing}
\end{algorithm}

\subsection{Initial Corpus via $\mu_c$}\label{sec:initial-corpus}

$\textsc{InitialCorpus}(S)$ applies every $\mu_c$ (\autoref{sec:semantic-model}) to every seed in $S$ (the \icode{riscv-tests} suite, \autoref{sec:implementation}), placing inputs that already exercise each of the 60 obligations into the corpus before iteration~1. \icode{riscv-tests} programs do not exercise immediates at byte boundaries, traces past $2^k$ thresholds, or similar obligation-critical conditions; random mutations rarely produce them either. $\mu_c$ shifts these conditions from ``must be discovered'' to ``available before iteration~1,'' letting the bandit spend its budget on combinatorial mutations of obligation-active inputs.

\subsection{Bandit Reward via $\kappa_c$}\label{sec:signature-feedback}

\tool replaces edge coverage (e.g., AFL~\cite{afl}), which targets executor branches uncorrelated with constraint structure, with \emph{signature novelty} computed from $\{\kappa_c\}_{c \in \mathcal{C}}$: an input is added to the corpus iff its signature $\sigma(x)$ (\autoref{eq:signature}) is unseen. $\kappa_c$ projects each trace into a 60-dimensional constraint-coverage vector that captures exactly the structure where bugs hide. The bandit reward in \autoref{alg:fuzzing} is fine-grained:
\[
\textsc{Reward}(\sigma, \mathit{Seen}) \;=\; \mathbf{1}\bigl[\sigma \notin \mathit{Seen}\bigr] \,+\, 0.25 \cdot \bigl|\sigma \setminus \mathit{SeenObls}\bigr|,
\]
where $\mathit{SeenObls} = \bigcup_{\sigma' \in \mathit{Seen}} \sigma'$ aggregates the individual obligation IDs activated by any signature observed so far. The first term rewards novel signature \emph{combinations} (weight~$1.0$); the second provides a gentler signal for inputs that activate previously unseen \emph{individual} obligations (weight~$0.25$ each).

\subsection{ISA-Aware Bandit-Guided Mutation}\label{sec:bandit-mutation}

$\textsc{Mutate}(\mathit{corpus}, \mathcal{B})$ chooses an input from the corpus, selects a mutation arm via the bandit~$\mathcal{B}$, applies the corresponding ISA-aware mutation, and returns the new input together with the chosen arm.

\paragraph{Mutation strategies}
\tool employs 8 mutation strategies, each operating at the RISC-V instruction level rather than the byte level (\autoref{tbl:mutation-arms}). Unlike generic byte-level mutations (e.g., bit flips, byte swaps), these strategies understand the RV32IM encoding format: they decode instructions into fields (opcode, rd, rs1, rs2, funct3, funct7, immediate), mutate semantically meaningful components, and re-encode valid instructions. This ISA awareness is critical because random byte-level mutations on 32-bit instruction words almost never produce valid RISC-V instructions, let alone programs that exercise interesting constraint behaviors. The ISA-aware encoder/decoder is validated against the official \icode{riscv-tests} suite (all RV32IM encodings round-trip correctly).

\begin{table}[t]
\centering
\resizebox{\columnwidth}{!}{%
\small
\begin{tabular}{clp{4.8cm}}
\toprule
\textbf{Arm} & \textbf{Strategy} & \textbf{Description} \\
\midrule
0 & Splice & Cross two corpus entries at a random point \\
1 & Register swap & Replace rd/rs1/rs2 with a register already used in the program \\
2 & Constant mutation & Replace immediates with boundary values \\
3 & Insert instruction & Append a random instruction; prefer memory ops when address-producing registers exist \\
4 & Delete instruction & Remove one instruction from the sequence \\
5 & Duplicate instruction & Copy an instruction to increase data locality \\
6 & Swap adjacent & Exchange two neighboring instructions \\
7 & Mnemonic replace & Substitute with a same-format mnemonic (e.g., \icode{add}$\leftrightarrow$\icode{sub}, \icode{lw}$\leftrightarrow$\icode{lh}) \\
\bottomrule
\end{tabular}%
}
\caption{ISA-aware mutation strategies (bandit arms). Arms 1, 2, 3, and 7 embed obligation awareness.}
\label{tbl:mutation-arms}
\end{table}

\paragraph{Obligation-arm alignment}
Obligations influence mutation through two complementary mechanisms: \emph{static design} embeds domain knowledge into specific arms, while \emph{dynamic adaptation} lets the bandit learn which arms are productive per backend.

Four arms encode direct obligation awareness. Arm~1 (register swap) can redirect \icode{rd} to \icode{x0}, directly triggering the \icode{decode.zero\_reg} obligation (the mechanism behind one of the x0 zero-days), and can set \icode{rs1}$=$\icode{rs2}, triggering \icode{exec.operand\_bind}. Arm~2 (constant mutation) replaces immediates with boundary values that stress both byte-piece decomposition (ALU, Arithmetic obligations) and $2^k$-threshold transitions (Memory, Execution, Lookup obligations). Arm~3 (insert instruction) biases toward memory operations when address-producing registers exist, driving memory footprint toward $2^k$ capacity thresholds (Memory, Row obligations). Arm~7 (mnemonic replace) tests opcode-selector constraints by substituting same-format instructions (e.g., \icode{add}$\leftrightarrow$\icode{sub} for ALU obligations, \icode{beq}$\leftrightarrow$\icode{blt} for \icode{control.branch\_cond}, \icode{lw}$\leftrightarrow$\icode{lh} for Memory sign-extension obligations).

The remaining four arms (0: splice, 4: delete, 5: duplicate, 6: swap adjacent) are structurally generic: they create input diversity without targeting any specific obligation. Their value is that they produce novel instruction sequences that the obligation-aware arms alone would not reach.

After execution, \tool computes the obligation signature and assigns a novelty-based reward to the selected mutation arm (\autoref{sec:signature-feedback}). The bandit uses this feedback to adapt subsequent arm selection to each backend's observed obligation coverage.

\paragraph{UCB1 selection}
\tool uses a UCB1 variant~\cite{ucb1} with a tunable exploration constant and an $\varepsilon$-greedy override. At each iteration, with probability $1-\varepsilon$ the arm with the highest upper confidence bound is selected,
\[
a^* = \arg\max_{a \in \{0,\dots,7\}} \left( \bar{r}_a + \alpha \sqrt{\frac{\ln n}{n_a}} \right),
\]
where $\bar{r}_a$ is arm $a$'s mean reward, $n_a$ is its pull count, $n$ is the total pull count, and $\alpha = 1.5$ is the exploration constant; with probability $\varepsilon = 0.05$, a uniformly random arm is selected instead, ensuring all arms continue to be exercised after UCB1's confidence bounds have converged. The $\alpha$-tuning and $\varepsilon$-override depart from textbook UCB1 (which fixes $\alpha = \sqrt{2}$ and uses no $\varepsilon$); both adjustments were retained because they consistently improved time-to-finding on a held-out tuning set disjoint from the 37-bug evaluation benchmark.

\subsection{Witness Probe via $\iota_c$}\label{sec:witness-search}

Underconstraint bugs do not manifest as register mismatches: they allow the prover to substitute an \emph{alternative witness} that the circuit still accepts. $\textsc{WitnessProbe}(C, B)$ detects them using each activated obligation's injection template $\iota_c$, which makes the search feasible by anchoring it at the obligation-active step rather than across the whole trace.

\paragraph{Procedure}
$\textit{injection\_targets}(C)$ maps activated obligations to $(c, s)$ anchor pairs. For each pair, $\iota_c$ generates candidate modifications $w'$ at trace steps within a window $[s - 16,\, s + 64]$ around the anchor, capped at $T_{\max} = 64$ trials per obligation. For every candidate with $w' \neq w_s$, \tool invokes the backend's $\textsc{constraint\_check}(w')$ function and flags the input as an \emph{underconstrained candidate} if the constraint system accepts. \textsc{WitnessProbe} returns the set of such flagged inputs.

\paragraph{Direct algebraic check}
Each backend exposes a $\textsc{constraint\_check}$ function that evaluates its constraint polynomials (AIR/RAP constraints in STARK-based backends) on the (possibly modified) trace and returns whether all constraint polynomials evaluate to zero. The check covers both \emph{boundary constraints} (single-row conditions such as range checks and boolean constraints) and \emph{transition constraints} (adjacent-row conditions such as register-file consistency and memory timestamp ordering) by supplying the modified row along with its neighbors. This is a \emph{direct algebraic check} on the constraint system: it does not require generating a full proof (which would take minutes), only evaluating the local constraint window (which takes microseconds). Cross-table constraints enforced via lookup or permutation arguments (e.g., Lasso or LogUp in various backends) are \emph{not} checked, since verifying these requires a global accumulator pass over the entire trace; the rare false-positive candidates this admits are eliminated by the full-proof validation below. The injection type is determined by the obligation's functional group: flipping a byte-piece value for ALU obligations, altering an address for memory obligations, toggling a boolean for lookup obligations, or perturbing a $2^k$ boundary value for obligations at power-of-two table thresholds (e.g., memory capacity, segment boundary, range-table size).

\paragraph{Soundness classification and limitations}
An injection is a true positive if (1)~the modification is non-trivial ($w' \neq w$), (2)~the constraint system accepts $w'$, and (3)~$w'$ encodes a computation different from RISC-V semantics. Condition~(3) is satisfied by design: each injection type produces a semantically different result (flipped byte-pieces change integer values, toggled booleans change control flow). Rare semantically equivalent alternatives are filtered by post-injection output comparison. The local check is only a fast filter during fuzzing: every flagged candidate is replayed with a prover that embeds the injected witness, a complete proof is generated, and the zkVM's unmodified verifier is run; an underconstraint is reported only if the verifier accepts.

The patterns are heuristic single-field perturbations (byte-piece flips, address alterations, boolean toggles, $2^n$ boundary shifts) at individual constraint steps. The 12 missed known bugs require cross-component reasoning beyond these single-field patterns (\autoref{sec:eval-rq1}). Injection tests only local effects; propagated effects are caught only if downstream divergence manifests in the differential detector or triggers a separate injection.

\subsection{Differential Detector}\label{sec:differential-detector}

The differential detector compares the 32 architectural registers returned by $\textsc{OracleRun}$ and $\textsc{BackendRun}$: any input with $r_O \neq r_B$ is flagged as a divergence bug. Register comparison detects every divergence gap whose deviation propagates to the final register state, by construction. Crashes are routed through the assertion-bypass instrumentation (\autoref{sec:instrumentation}), which converts fatal Rust assertions to logged warnings so the executor continues to a (corrupted) final state that remains comparable to the oracle's. We compare registers rather than full memory to avoid normalizing heterogeneous address-space layouts; memory-only divergences are partially addressed by witness injection at memory-related constraint steps.

\subsection{Reference Oracle}\label{sec:oracle}

$\textsc{OracleRun}(x', O)$ executes $x'$ on a reference RV32IM simulator built on \icode{rrs-lib}, validated against the official \icode{riscv-tests} compliance suite, and returns the post-execution architectural register state $r_O$. It supports two memory models (\icode{SharedCodeData} and \icode{SplitCodeData}), selected per backend to match the backend's architecture and avoid false positives. Edge cases follow the RISC-V specification: \icode{ECALL} is a no-op recording arguments, \icode{EBREAK} traps, \icode{FENCE} is a no-op, and misaligned accesses trap (flagged as a divergence bug if the backend handles them silently). The oracle also pre-filters inputs exceeding a configurable step limit.

$\textsc{BackendRun}(x', B)$ executes $x'$ on the instrumented backend (\autoref{sec:instrumentation}), returning the same register state $r_B$ together with the list $C$ of obligation activations evaluated during trace generation.

\subsection{Implementation}\label{sec:implementation}

\tool is approximately 12{,}000 lines of Rust (backend instrumentation per \autoref{sec:instrumentation}, obligation evaluation, witness injection, ISA-aware mutation) and 3{,}000 lines of Python (RL loop, bandit scheduling, corpus management). The base seed set is 2{,}172 \icode{riscv-tests} sequences (32-bit instruction words, up to 27 instructions per seed; mutated inputs capped at 256 instructions); the seeds cover all RV32IM operations including edge cases (division by zero, signed overflow, misaligned memory). Proof generation is skipped during fuzzing (flagged candidates are validated with complete proofs, \autoref{sec:witness-search}): each iteration evaluates only trace generation plus the local constraint-check (AIR/RAP polynomials on the modified row and its neighbors), reducing per-input cost from minutes to milliseconds.

\section{Evaluation}\label{sec:eval}

We evaluate \tool to answer the following research questions:
\begin{itemize}
    \item \textbf{RQ1 (Effectiveness):} How effective is \tool at discovering semantic gaps (divergence and underconstraint) compared to state-of-the-art zkVM testing approaches?
    \item \textbf{RQ2 (Ablation):} What is the contribution of each obligation hook: seed enrichment, obligation-signature reward, and witness injection?
    \item \textbf{RQ3 (Zero-Days):} Can \tool discover previously unknown vulnerabilities in production-grade zkVMs?
\end{itemize}

\subsection{Experimental Setup}\label{sec:eval-setup}

\paragraph{Target zkVMs}
We evaluate \tool on six production-grade RISC-V zkVMs spanning four proof architectures (STARK/Plonky3, STARK/FRI, Lookup/Lasso, Nova/IVC): SP1, Pico, OpenVM, Risc0, Jolt, and Nexus (\autoref{tbl:backends}). For each system, we test against one or more pinned commit snapshots to ensure reproducibility.

\paragraph{Seed corpus and obligation enrichment}
The base seed set is 2{,}172 \icode{riscv-tests} sequences (\autoref{sec:implementation}); $\mu_c$-enrichment then constructs the loop's initial corpus (\autoref{sec:initial-corpus}).

\paragraph{Baselines (for RQ1)}
RQ1 compares \tool against two external baselines; the ablation variants of \tool itself are introduced separately in RQ2 (\autoref{sec:eval-rq2}).
\begin{itemize}
    \item \textbf{\arguzz}~\cite{arguzz}: a state-of-the-art dynamic testing framework for zkVMs that combines metamorphic testing with fault injection. We re-ran \arguzz from its public artifact on the same hardware, with the same per-backend time and core budget as \tool, against the same six backends and the same 37-bug benchmark.
    \item \textbf{Baseline}: a generic fuzzer with uniform random byte-level mutation and differential register comparison only; it lacks semantic feedback, obligation guidance, ISA-aware operators, and witness injection.
\end{itemize}

\paragraph{Configuration}
All experiments are conducted on an AMD EPYC 9654 server with 96 cores / 192 hardware threads, 768~GB RAM, and 2$\times$4~TB NVMe storage, running Ubuntu 22.04. Each (tool, backend) configuration is evaluated in five independent 24-hour campaigns with distinct random seeds, for a total budget of 120 hours per configuration. To characterize variability due to bandit scheduling and randomized mutations, we report the union of distinct bugs detected across runs, the distribution of per-run detection counts, and per-bug wall-clock discovery times. \tool uses UCB1 with exploration constant $\alpha = 1.5$ and an $\varepsilon$-greedy override at $\varepsilon = 0.05$, an asymmetric witness-injection search window extending 16 steps before and 64 steps after the anchor, and a maximum of 64 injection trials per obligation.

\paragraph{Metrics}
We report \textbf{unique bugs}, the number of distinct divergence and underconstraint bugs detected.

\subsection{RQ1: Bug Detection Effectiveness}\label{sec:eval-rq1}

\begin{figure}[t]
\centering
\begin{tikzpicture}
\begin{axis}[
    ybar=1.5pt,
    bar width=6pt,
    width=\columnwidth,
    height=0.55\columnwidth,
    ymin=0, ymax=12,
    ylabel={Bugs detected},
    ylabel style={font=\small},
    symbolic x coords={SP1, Pico, Risc0, Jolt, Nexus, OpenVM},
    xtick=data,
    x tick label style={font=\footnotesize},
    y tick label style={font=\footnotesize},
    nodes near coords,
    nodes near coords style={font=\scriptsize, anchor=south},
    enlarge x limits=0.10,
    ymajorgrids=true,
    grid style={dashed, gray!40},
    legend style={
        at={(0.5,1.02)}, anchor=south,
        legend columns=3, font=\footnotesize,
        draw=black!30, /tikz/every even column/.append style={column sep=0.4cm},
    },
]
\addplot[draw=black, fill=black!75]
    coordinates {(SP1,2) (Pico,1) (Risc0,1) (Jolt,7) (Nexus,3) (OpenVM,11)};
\addplot[draw=black, fill=black!40]
    coordinates {(SP1,1) (Pico,0) (Risc0,0) (Jolt,3) (Nexus,1) (OpenVM,0)};
\addplot[draw=black, fill=black!10]
    coordinates {(SP1,0) (Pico,0) (Risc0,0) (Jolt,2) (Nexus,0) (OpenVM,1)};
\legend{\tool, \arguzz, Baseline}
\end{axis}
\end{tikzpicture}
\caption{Known-bug detections per backend (totals: SP1~9, Pico~2, Risc0~5, Jolt~7, Nexus~3, OpenVM~11). \tool bars report the union over five runs.}
\label{fig:effectiveness}
\end{figure}
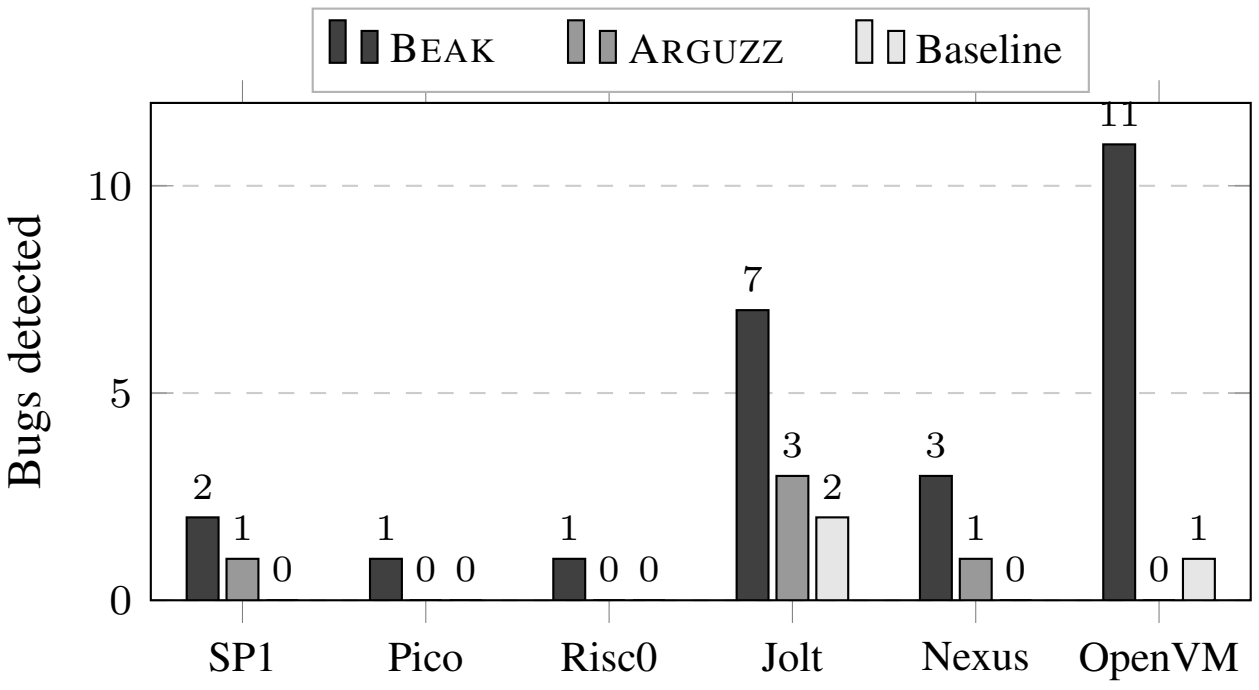

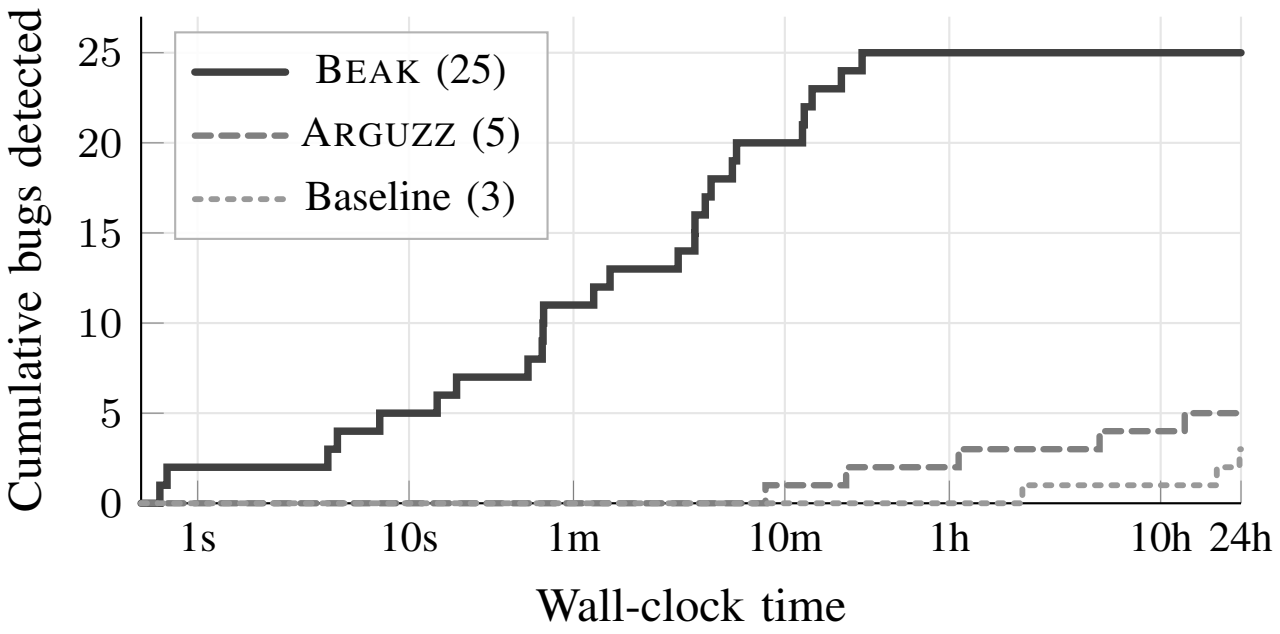
\begin{figure}[t]
    \centering
    \begin{tikzpicture}
    \begin{axis}[
        width=\columnwidth,
        height=4.8cm,
        xlabel={Wall-clock time},
        ylabel={Cumulative bugs detected},
        ylabel style={yshift=-2pt},
        xmode=log,
        xmin=0.00015, xmax=24,
        ymin=0, ymax=27,
        xtick={0.000277777778,0.002777777778,0.016666666667,0.166666666667,1,10,24},
        xticklabels={1s,10s,1m,10m,1h,10h,24h},
        ytick={0,5,10,15,20,25},
        legend style={
            at={(0.03,0.97)},
            anchor=north west,
            font=\footnotesize,
            draw=black!30,
            fill=white,
            fill opacity=0.9,
            text opacity=1,
            row sep=1pt,
        },
        grid=major,
        grid style={gray!20},
        tick label style={font=\footnotesize},
        label style={font=\small},
        axis y line*=left,
        axis x line*=bottom,
        clip=false,
    ]
    \addplot[black!75, line width=1.4pt, mark=none, const plot mark left, line cap=round]
        coordinates {
            (0.00015,0) (0.000184,1) (0.000199,2) (0.001147,3) (0.001273,4)
            (0.002019,5) (0.003775,6) (0.004663,7) (0.010155,8) (0.011854,9)
            (0.011963,10) (0.012036,11) (0.020771,12) (0.024823,13) (0.052149,14)
            (0.062531,15) (0.062614,16) (0.070059,17) (0.074627,18) (0.094122,19)
            (0.098471,20) (0.202014,21) (0.205960,22) (0.224296,23) (0.308593,24)
            (0.386732,25) (24,25)
        };
    \addlegendentry{\textsc{Beak} (25)}
    \addplot[black!50, line width=1.2pt, mark=none, const plot mark left, line cap=round, densely dashed]
        coordinates {
            (0.00015,0) (0.135,1) (0.325,2) (1.107,3) (5.141,4) (13,5) (24,5)
        };
    \addlegendentry{\textsc{Arguzz} (5)}
    \addplot[black!40, line width=1.1pt, mark=none, const plot mark left, line join=round, line cap=round, dotted]
        coordinates {
            (0.00015,0) (2.217,1) (18.433,2) (23.617,3) (24,3)
        };
    \addlegendentry{Baseline (3)}
    \end{axis}
    \end{tikzpicture}
    \caption{Cumulative detection of known bugs. \tool uses each bug's earliest detection over five runs; \arguzz and Baseline show their best campaign.}
    \label{fig:cumulative-bugs}
\end{figure}

To assess \tool's effectiveness, we construct a ground-truth benchmark of 37 independently sourced known zkVM semantic-gap bugs, separate from the 14 zero-day findings discovered by \tool. The known benchmark is drawn from three sources: 11 bugs previously reported by \arguzz~\cite{arguzz}, 25 bugs from professional security audits of production zkVMs~\cite{audit_sp1_cantina, audit_sp1_kalos, audit_sp1_rkm0959, audit_sp1_veridise, audit_sp1_v4, audit_openvm_cantina, audit_openvm_internal, audit_risc0_hexens, audit_risc0_veridise}, and 8 bugs from public issue trackers (with overlap: some bugs appear in multiple sources). The benchmark partitions cleanly along the divergence/underconstraint axis introduced in \autoref{sec:formal-properties}: 28 are underconstraints (the executor computes the correct state but the constraint system admits an alternative witness), and 9 are divergences (the executor itself deviates from the ISA on a legitimate input). For each bug, we determine whether each tool can detect it through the corresponding detector: register-level comparison for divergence, and witness injection for underconstraint---an underconstraint counts as detected only when the mutated witness is carried through complete proof generation and the resulting proof is accepted by the zkVM's own verifier. \autoref{fig:effectiveness} presents per-backend detections; time-to-finding dynamics are reported separately in \autoref{fig:cumulative-bugs}.

On the 37 known bugs, \tool detects 25/37 (67.6\%), compared with 5/37 (13.5\%) for \arguzz and 3/37 (8.1\%) for Baseline. The detected set includes all bugs found by either baseline. These detections comprise all 9 divergence bugs in the benchmark and 16 of the 28 underconstraint bugs. The divergence cases manifest as register mismatches or crashes during non-injected execution, whereas the underconstraint cases are detected through obligation-guided injection. The results illustrate the complementary coverage of execution-side failure detection and verifier-side underconstraint detection.

Underconstraint validation combines evidence of an obligation violation (\autoref{sec:formal-properties}) with complete proof generation for the mutated execution and acceptance by the target zkVM's verifier. All reported underconstraints satisfied this validation criterion.

\paragraph{Per-backend analysis}
\tool detects all known benchmark bugs in Jolt (7/7), Nexus (3/3), and OpenVM (11/11), but achieves lower coverage in SP1 (2/9), Pico (1/2), and Risc0 (1/5). Overall, 12 of the 37 benchmark bugs remain undetected.

All 12 missed bugs require cross-component reasoning beyond single-step obligations. Five require coordinated modifications to multiple witness fields, such as an address tuple and its auxiliary range-check cells. The other seven involve backend-specific structures (e.g., SP1's recursion VM) that the common obligation set does not cover. These limitations motivate dependency-guided joint mutation and additional backend-specific predicates, enrichment templates, and injection templates. Evaluating these extensions is left to future work.

\begin{mdframed}[
    backgroundcolor=gray!10,
    linecolor=white,
    skipabove=0.5em
    ]\noindent
    \textbf{Result for RQ1:} Across five independent runs, \tool detects 25/37 known bugs (67.6\%), compared with 5/37 (13.5\%) for \arguzz and 3/37 (8.1\%) for Baseline. Its detected set includes all bugs found by either baseline and contains $5.0\times$ as many bugs as \arguzz's. The mean per-run detection count is 22 (95\% confidence interval: $[19.1, 24.9]$). All reported underconstraints were validated by complete proof generation and verifier acceptance.
\end{mdframed}

\begin{figure*}[!t]
    \centering
    \input{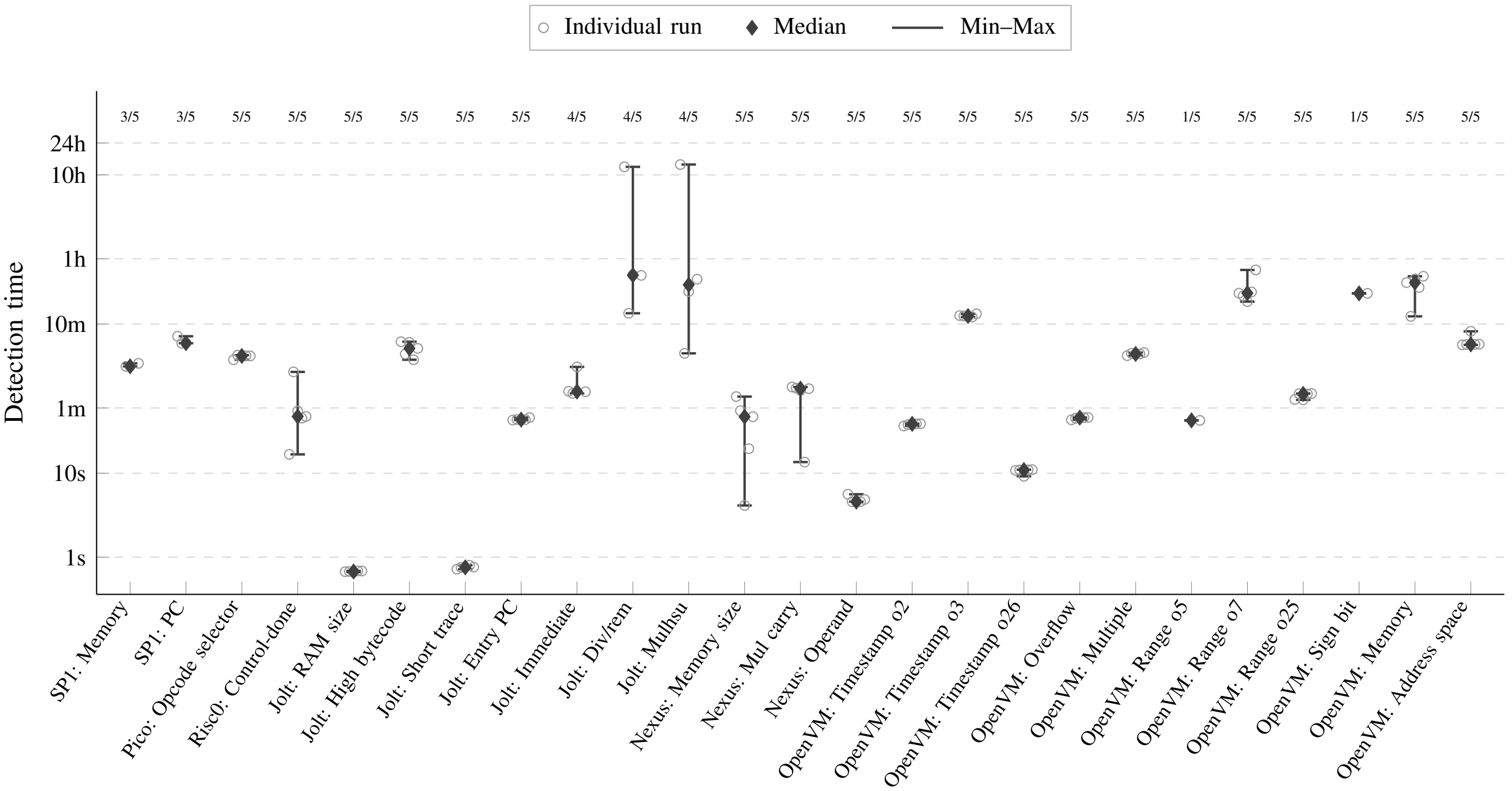}
    \caption{Detection time for each of the 25 known bugs \tool detects, across five independent runs. Circles mark individual runs, diamonds the median, and bars the min--max range; the label above each bug gives how many of the five runs detected it.}
    \label{fig:time-to-bug}
\end{figure*}

\paragraph{Temporal dynamics and run-to-run variance}
\autoref{fig:cumulative-bugs} summarizes discovery times on the 37-bug benchmark. Aggregating the earliest detection of each bug across five independent runs, \tool reaches 21 distinct bugs by 0.21~hours and all 25 detected bugs by 0.39~hours. \autoref{fig:time-to-bug} complements this aggregate view with per-bug discovery times and detection frequencies across the five runs. For each bug, we compute the median discovery time over runs in which it was detected. These per-bug medians range from 0.7~seconds to 0.64~hours, with a median of 0.03~hours (approximately two minutes). The latest observed detection in an individual run occurs at 13.3~hours.

The five runs detect 22, 23, 23, 18, and 24 distinct bugs, respectively (median 23; range 18--24). Of the 25 bugs detected across runs, 18 are detected in all five runs, three in four runs, two in three runs, and two in one run each. Thus, 18 of the 25 bugs are consistently detected across all runs, while detection of the remaining seven varies across runs.

In the baseline campaigns shown, \arguzz detects five bugs by 13~hours, and Baseline detects three by approximately 23.6~hours.

\subsection{RQ2: Ablation Study}\label{sec:eval-rq2}

\begin{figure}[t]
\centering
\begin{tikzpicture}
\begin{axis}[
    ybar,
    bar width=22pt,
    width=\columnwidth,
    height=0.55\columnwidth,
    ymin=0, ymax=30,
    ylabel={Bugs detected (of 37)},
    ylabel style={font=\small},
    xtick={0,1,2,3},
    xticklabels={\tool, \tool$^{*}$, \tool$^{\dagger}$, \tool$^{\ddagger}$},
    x tick label style={font=\footnotesize},
    y tick label style={font=\footnotesize},
    nodes near coords,
    nodes near coords style={font=\small, anchor=south},
    enlarge x limits=0.12,
    ymajorgrids=true,
    grid style={dashed, gray!40},
]
\addplot[ybar, bar shift=0pt, draw=black, fill=black!75]
    coordinates {(0,25)};
\addplot[ybar, bar shift=0pt, draw=black, fill=black!50]
    coordinates {(1,16)};
\addplot[ybar, bar shift=0pt, draw=black, fill=black!50]
    coordinates {(2,19)};
\addplot[ybar, bar shift=0pt, draw=black, fill=black!25]
    coordinates {(3,9)};
\end{axis}
\end{tikzpicture}
\caption{Ablation on the 37 known bugs (120 hours per backend). $^{*}$ removes $\mu_c$; $^{\dagger}$ replaces $\kappa_c$ with edge coverage; $^{\ddagger}$ removes all three hooks.}
\label{fig:ablation}
\end{figure}
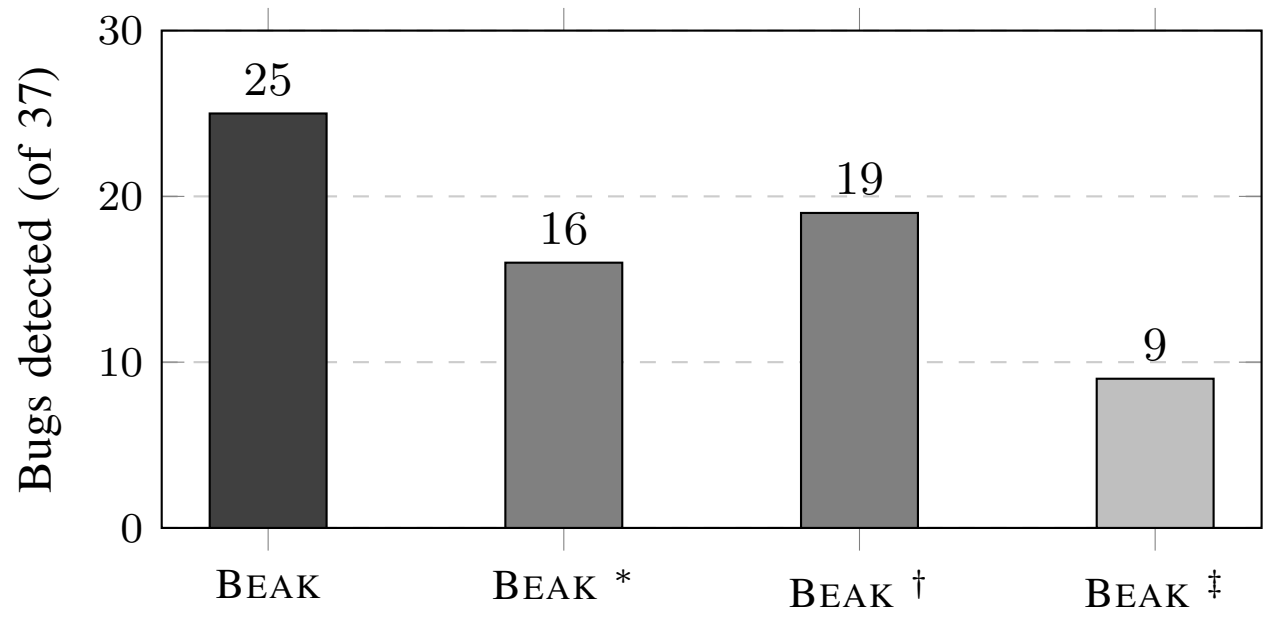

We ablate \tool's obligation hooks by replacing each with the default-random mechanism it displaces. All ablation variants below retain \tool's ISA-aware mutation operators (\autoref{tbl:mutation-arms}); they differ only in which obligation hook is removed.
\begin{itemize}
    \item \textbf{\tool$^{*}$}: removes specification-driven enrichment $\mu_c$ and starts from raw \icode{riscv-tests}.
    \item \textbf{\tool$^{\dagger}$}: replaces obligation-signature novelty $\kappa_c$ with edge-coverage reward.
    \item \textbf{\tool$^{\ddagger}$}: removes all three hooks (raw seeds, edge-coverage reward, blind unanchored injection); the all-hooks-off reference point.
\end{itemize}
\autoref{fig:ablation} reports detections on the 37 known bugs under the same 120-hour-per-backend budget as RQ1.

\paragraph{Effect of obligation guidance}
The ablations show that the three hooks contribute complementary signal. Removing $\mu_c$ drops detections from 25/37 to 16/37: the raw seed corpus exercises legal ISA behavior, but only sparsely reaches obligation-critical conditions such as byte-boundary immediates or $2^n$ table thresholds. Replacing $\kappa_c$ with edge coverage yields 19/37, indicating that executor-state novelty is an imprecise proxy for constraint-state novelty.

\paragraph{Effect of anchored injection}
The injection hook is not meaningfully isolated by a single-hook blind variant: production traces contain millions of witness cells, whereas a bug-relevant obligation neighborhood spans only a small local window. This search asymmetry makes blind witness-cell injection infeasible at the evaluated budget. Consistent with this structural argument, \tool$^{\ddagger}$, which removes all obligation guidance and uses blind unanchored injection, detects only 9/37, and the full pipeline detects nearly three times as many bugs as this no-obligation variant, reaching 25/37. These 9 are exactly the benchmark's divergence bugs, found by the differential detector; blind injection detects none of the 28 underconstraints. An injection-only variant retaining $\iota_c$ but replacing $\mu_c$ and $\kappa_c$ with defaults would in principle isolate injection's contribution; without $\kappa_c$ to identify active obligations, $\iota_c$ has no anchor points and degrades to blind injection.

\begin{mdframed}[
    backgroundcolor=gray!10,
    linecolor=white,
    skipabove=0.5em
    ]\noindent
    \textbf{Result for RQ2:} Each obligation hook is load-bearing. Removing seed enrichment, obligation-signature reward, or all hooks reduces detection to 16/37, 19/37, and 9/37, respectively; the full pipeline reaches 25/37.
\end{mdframed}

\subsection{RQ3: Zero-Day Discovery}\label{sec:eval-rq3}

We applied \tool to the latest development branches of six zkVM systems and discovered 14 zero-day findings. These include critical soundness violations enabling proof forgery, executor trace-divergence bugs, and production-crashing failures. All were responsibly disclosed, confirmed, and patched; Appendix~\ref{app:disclosure} lists each finding.

\paragraph{Case study: x0 register overwrite}
The same vulnerability class exists independently in three backends: one allowed an ecall to write to x0; another lacked a circuit-layer zero-register constraint; a third aliased registers with memory, so \icode{sw} to address~0 overwrites x0. All three map to the \icode{decode.zero\_reg} obligation. \tool detects all three instances automatically through its standard fuzzing loop, and each was confirmed by the respective maintainers.

\paragraph{Case study: \icode{is\_real} bypass and timestamp wrap}
In one backend, the \icode{MemoryLocalChip} omits the boolean constraint on the \icode{is\_real} lookup multiplicity; the \icode{lookup.boolean\_mult} obligation anchors injection of a non-boolean value, and the circuit accepts. In the same backend, the timestamp monotonicity check wraps modulo $p \approx 2^{31}$; the \icode{time.fp\_wrap} obligation anchors injection of a near-$p$ timestamp. Both are underconstraints, invisible to differential testing; as in RQ1, each underconstraint zero-day was validated by complete proof generation and verifier acceptance.

\paragraph{Security impact}
The critical underconstraint zero-days enable forging proofs for incorrect RISC-V execution. For example, in a rollup whose state transitions are proved by an affected zkVM, a forged proof of a withdrawal that never executed would be accepted on-chain, enabling unauthorized withdrawals or unbounded minting. We did not construct end-to-end exploits against deployed systems. The x0 vulnerability is particularly noteworthy: the same ISA-level invariant was independently violated in three unrelated codebases.

\begin{mdframed}[
    backgroundcolor=gray!10,
    linecolor=white,
    skipabove=0.5em
    ]\noindent
    \textbf{Result for RQ3:} 14 zero-day findings across four of the six backends, all confirmed and patched. Three share a common semantic class (x0 overwrite), demonstrating cross-backend generality of the semantic model.
\end{mdframed}

\section{Related Work}\label{sec:related}

\paragraph{Zero-knowledge proof security}
\zkap~\cite{zkap} applies static analysis to Circom circuits; Picus~\cite{picus} verifies Circom witness uniqueness via SMT; zkFuzz~\cite{zkfuzz} (IEEE S\&P~2026) uses mutation-based dynamic analysis for Circom circuits, combining trace characterization with joint program/input mutation and an error-based fitness signal; Coda~\cite{coda} verifies ZK circuit implementations against high-level specifications via refinement types and Coq-backed proof obligations. All four target DSL-level circuits or circuit programs rather than instruction-set-level zkVM semantics. CircUZZ~\cite{circuzz} (CCS~2025) and MTZK~\cite{mtzk} (NDSS~2025) apply metamorphic testing to ZK \emph{compiler pipelines}: CircUZZ tests four pipelines (Circom, Corset, Gnark, Noir) by stacking equivalence-preserving circuit rewrites, while MTZK designs metamorphic relations targeting ZK compiler correctness. Both operate at the DSL-to-circuit compilation boundary and detect compiler-introduced bugs; neither addresses zkVM-level constraint soundness or RISC-V semantic fidelity. zkVM arithmetizations differ in two ways: \emph{scale} ($10^5$--$10^6$ AIR/lookup constraints with cross-linked trace tables over small-characteristic fields such as BabyBear, KoalaBear, and Goldilocks, beyond SMT solver capacity in practice) and \emph{specification} (verifying soundness against underconstraint requires modeling both the constraint system and RISC-V semantics; we are unaware of a production zkVM that exposes such a specification). FORAY~\cite{foray} synthesizes DeFi attacks at the application level; Tabby~\cite{tabby} optimizes circuit \emph{efficiency} via program synthesis, complementary to \tool's focus on \emph{correctness}.

\arguzz~\cite{arguzz} is the closest work: it applies metamorphic testing with fault injection to zkVMs and targets both prover-side divergence and verifier-side underconstraint gaps. Unlike \tool, \arguzz does not use an obligation model to couple input generation with injection-site selection. In the motivating example of \autoref{sec:motivating-example}, \tool constructs an input with immediate~$-1$ through $\mu_c$ and uses the corresponding $\kappa_c$ activation to anchor $\iota_c$ at the ALU limb-decomposition step. This explicitly links the input's trigger condition to the witness location being tested. \zkap~\cite{zkap} instead performs static analysis of Circom circuits, rather than analyzing zkVM execution traces. \tool's design adds two structural pieces: (1)~the obligation triple $(\mu_c, \kappa_c, \iota_c)$ restructures the three default-random decision points of the fuzzing loop (initial corpus, RL reward, witness probe) so that constraint corner cases are exercised systematically, and (2)~witness injection anchored at $\kappa_c$-active steps via $\iota_c$ reduces the per-obligation probe to a tractable per-step check. On the known benchmark, \tool detects 25/37 to \arguzz's 5/37 ($5.0\times$) under the same time budget, and \tool's detections include all of \arguzz's. The additional bugs are predominantly underconstraints, consistent with the structural advantage of obligation-anchored injection.

\paragraph{Domain-specific coverage metrics}
\tool belongs to a growing trend replacing edge coverage with domain-specific metrics. Data Coverage~\cite{datacoverage} (USENIX Security 2024) tracks constant-data references as fuzzing feedback, exposing behavior that code coverage alone can miss. zkFuzz extends this idea inside individual Circom circuits via trace characterization and error-guided mutation; \tool extends it from DSL-level circuits to RV32IM zkVMs along two axes: (1)~obligations form a triple that simultaneously seeds the corpus ($\mu_c$), defines the reward ($\kappa_c$), and anchors witness injection ($\iota_c$), so a single specification artifact drives three loop decisions instead of feeding a single coverage metric; and (2)~the reward combines novel obligation-signature \emph{combinations} with a per-obligation bonus for newly activated obligations (\autoref{sec:signature-feedback}), giving more sustained signal than flat set-based coverage in the large but repetitive constraint spaces of zkVMs.

\paragraph{Differential testing, fuzzing, and mutation scheduling}
Differential testing~\cite{mckeeman1998differential,csmith} compares outputs of multiple implementations; \tool adapts this to zkVMs with obligation-signature novelty feedback. Coverage-guided fuzzers (AFL~\cite{afl}, LibFuzzer~\cite{libfuzzer}) use edge coverage, inadequate for zkVMs where interesting behaviors are at the constraint level; Klees et al.~\cite{klees2018evaluating} further show that sound empirical comparison requires controlled statistical methodology, such as repeated runs. Grammar-aware fuzzers (Nautilus~\cite{nautilus}, Gramatron~\cite{gramatron}) ensure structural validity; \tool's ISA-aware mutations additionally incorporate constraint-obligation awareness. MOpt~\cite{mopt} and EcoFuzz~\cite{ecofuzz} apply optimization-based scheduling to mutation operators; NEZHA~\cite{nezha} uses $\delta$-diversity for differential testing; SemFuzz~\cite{semfuzz} extracts semantic information from CVE descriptions. Unlike these tools, in \tool the same obligation triple simultaneously seeds the corpus ($\mu_c$), defines the bandit reward via signature novelty ($\kappa_c$), and anchors witness injection ($\iota_c$); the three default-random decision points of a generic RL fuzzer are all replaced by obligation-driven structure.

\section{Conclusion}\label{sec:concl}

We presented \tool, a fuzzer that finds semantic gaps in RISC-V zkVMs, covering both soundness (underconstraint) and completeness (divergence) violations. \tool derives 60 constraint obligations from the ISA specification and the proof system's algebraic structure, and uses each obligation's triple $(\mu_c, \kappa_c, \iota_c)$ to replace the three random decisions of an RL fuzzer: the initial corpus, the reward, and where to inject into the witness. On 37 known bugs across six leading production zkVMs, \tool detects 25 versus \arguzz's 5, including all of \arguzz's, and it found 14 zero-day bugs in the latest development branches of these systems, all responsibly disclosed and patched.

Three lessons emerge. First, underconstraints outnumber divergences 28 to 9 in the known benchmark and are invisible to output comparison, so obligation-anchored witness injection is essential: with all three hooks removed, \tool detects only the 9 divergence bugs. Second, a single obligation model, derived from specifications rather than from observed bugs, applies uniformly across six backends spanning four proof architectures. Third, guidance matters beyond injection: removing $\mu_c$ or replacing $\kappa_c$ with edge coverage reduces detections to 16 and 19, respectively. The 12 missed bugs require cross-component reasoning; dependency-guided injection that jointly modifies related witness fields, and backend-specific obligations under the same interface, are the most promising next steps. The source code for \tool is publicly available at \url{https://github.com/whbjzzwjxq/beak-fuzz}.

\section*{Ethics Considerations}\label{sec:ethics}

This work uncovers soundness vulnerabilities in production zero-knowledge virtual machines (zkVMs) that secure billions of dollars in on-chain value. We took the following measures to ensure responsible conduct.

\paragraph{Responsible disclosure}
All 14 zero-day findings discovered by \tool were privately disclosed to the maintainers of the affected zkVMs through their established security channels (security email addresses or private bug-bounty platforms) before any public communication. We coordinated with each maintainer on triage, root-cause analysis, and remediation. Every finding has been confirmed by the maintainers and patched in the latest released versions of the affected systems prior to submission of this paper. We do not release any proof-of-concept exploit that could be weaponized against deployed systems.

\paragraph{Local-only experimentation}
All experiments were conducted on local instances of the open-source zkVM implementations using synthetic inputs. We did not interact with any production deployment, mainnet rollup, or live blockchain bridge. No real user funds, transactions, or private data were involved at any point.

\paragraph{Research artifacts}
The benchmark of known bugs we use is drawn from publicly disclosed audit reports and prior academic work; we do not redistribute non-public information. The source code and build and usage instructions for \tool are publicly available at \url{https://github.com/whbjzzwjxq/beak-fuzz}.

\section*{Acknowledgment}
We are grateful for the time and effort the anonymous reviewers invested in reviewing our work and offering feedback. This work is supported in part by an Ethereum Foundation Academic Award.

\bibliographystyle{IEEEtran}
\bibliography{ai}

\appendices
\section{Zero-Day Disclosure}\label{app:disclosure}

\autoref{tbl:disclosure} lists the 14 zero-day findings of \autoref{sec:eval-rq3}. All were privately disclosed, confirmed by the affected maintainers, and patched.

\begin{table}[!ht]
\centering
\footnotesize
\begin{tabular}{lll}
\toprule
\textbf{Finding} & \textbf{zkVM} & \textbf{Obligation} \\
\midrule
Reg-zero                     & Pico   & \texttt{decode.zero\_reg} \\
Address-alias      & Pico   & \texttt{mem.ts\_order} \\
JALR-odd-target        & Pico   & \texttt{control.jalr\_lsb} \\
IsReal-mult              & Pico   & \texttt{lookup.boolean\_mult} \\
Timestamp-wrap & Pico & \texttt{time.fp\_wrap} \\
Ecrecover-panic & Pico & \texttt{control.ecall\_arg} \\
Reg-zero                     & OpenVM & \texttt{decode.zero\_reg} \\
Poseidon2-empty  & OpenVM & \texttt{row.trace\_pow2} \\
FRI-zero-length          & OpenVM & \texttt{row.trace\_pow2} \\
Hintstore-zero     & OpenVM & \texttt{row.trace\_pow2} \\
Access-adapter      & OpenVM & \texttt{row.trace\_pow2} \\
Reg-zero                     & Risc0  & \texttt{decode.zero\_reg} \\
RW-memory & Nexus & \texttt{row.padding\_leak} \\
Register-memory & Nexus & \texttt{row.padding\_leak} \\
\bottomrule
\end{tabular}
\caption{Zero-day findings, all confirmed and patched.}
\label{tbl:disclosure}
\end{table}

\section{Full Obligation Enumeration and Detected Benchmark Bugs}\label{app:obligations}

This appendix lists all 60 constraint obligations enumerated by \tool, organized by the 10 functional groups of \autoref{tbl:obligations}. For each obligation, we give its identifier, a short description, and any linked bugs among the 25 detected by \tool on the benchmark of 37 known bugs (\autoref{sec:eval-rq1}). The detected set is the union across the five evaluation runs; undetected benchmark bugs are omitted. An $^{\mathrm{A}}$ superscript marks a bug also detected by \arguzz under the same time budget used in RQ1. An empty entry indicates that no detected benchmark bug is linked to that obligation. A single detected bug may map to multiple obligations (e.g., \texttt{Jolt-Dory-ShortTrace-01} maps to both \texttt{interaction.commit\_pow2} and \texttt{row.trace\_pow2}); the tables contain 25 distinct detected bugs.

\newcommand{\benchlist}[1]{\begin{tabular}[t]{@{}l@{}}#1\end{tabular}}

\begin{table*}[!t]
\centering
\footnotesize
\setlength{\tabcolsep}{4pt}
\begin{tabular}{p{4.3cm}p{5.7cm}p{6.6cm}}
\toprule
\textbf{Obligation} & \textbf{Description} & \textbf{Detected benchmark bug} \\
\midrule
\multicolumn{3}{l}{\emph{Decode (7)}} \\
\midrule
\texttt{decode.zero\_reg}            & x0 immutability: writes targeting \texttt{x0} must be discarded                &  \\
\texttt{decode.rd\_decomp}           & rd bit decomposition consistency over field encoding                           &  \\
\texttt{decode.upper\_imm}           & upper-immediate (LUI/AUIPC) materialization into 32-bit value                  & \benchlist{Jolt-Immediate-01$^{\mathrm{A}}$\\OpenVM-RangeCheck-Audit-o7} \\
\texttt{decode.operand\_route}       & operand routing from decoded fields to the execution pipeline                  &  \\
\texttt{decode.prog\_table\_pow2}    & program table $2^k$ capacity threshold                                         & \benchlist{Jolt-RamSize-01\\Jolt-HighBytecode-01} \\
\texttt{decode.fmt\_imm\_reasm}      & format-specific immediate reassembly and sign extension (I/S/B/U/J)            & OpenVM-SignBit-Audit-o8 \\
\texttt{decode.illegal\_opcode}      & illegal opcode trap                                                            &  \\
\midrule
\multicolumn{3}{l}{\emph{ALU (8)}} \\
\midrule
\texttt{alu.imm\_limb}               & immediate byte-piece (limb) decomposition and recomposition                    & OpenVM-RangeCheck-Audit-o5 \\
\texttt{alu.shift\_mask}             & shift-amount masking to low 5~bits                                             &  \\
\texttt{alu.cmp\_bool}               & SLT/SLTU comparison result is boolean                                          &  \\
\texttt{alu.add\_carry}              & ADD carry-chain consistency across limbs                                       &  \\
\texttt{alu.sub\_borrow}             & SUB borrow-chain consistency across limbs                                      &  \\
\texttt{alu.bitwise\_lookup}         & AND/OR/XOR lookup-table consistency                                            &  \\
\texttt{alu.cmp\_aux}                & comparison auxiliary witness consistency                                       &  \\
\texttt{alu.funct\_dispatch}         & funct3/funct7 dispatch to correct ALU operation                                &  \\
\midrule
\multicolumn{3}{l}{\emph{Arithmetic (6)}} \\
\midrule
\texttt{arith.div\_rem}              & division remainder bounds ($0 \le r < |\mathit{divisor}|$)                     & Jolt-DivRem-01 \\
\texttt{arith.signed\_overflow}      & signed overflow at $-2^{31}$ edge cases                                        & OpenVM-Multiple-Audit-o15 \\
\texttt{arith.mul\_decomp}           & multiplication product decomposition into hi/lo limbs                          & Nexus-MulCarry-01$^{\mathrm{A}}$ \\
\texttt{arith.div\_by\_zero}         & division-by-zero result convention ($q=2^{32}-1$, $r=\mathit{dividend}$)       &  \\
\texttt{arith.mulh\_consistency}     & MUL/MULH high-low consistency on the same operands                             &  \\
\texttt{arith.signed\_unsigned\_mul} & signed/unsigned multiplication sign correction                                 & Jolt-Mulhsu-01$^{\mathrm{A}}$ \\
\midrule
\multicolumn{3}{l}{\emph{Control (8)}} \\
\midrule
\texttt{control.pc\_limb}            & PC byte-piece consistency across steps                                         &  \\
\texttt{control.ecall\_arg}          & ecall argument register encoding                                               &  \\
\texttt{control.entrypoint\_bind}    & entrypoint PC binding at trace start                                           & Jolt-EntryPc-Binding-01 \\
\texttt{control.ecall\_word}         & ecall word validity (selector + argument range)                                &  \\
\texttt{control.branch\_cond}        & branch condition selection (signed vs.\ unsigned)                              &  \\
\texttt{control.branch\_target}      & branch target arithmetic (PC + sign-extended offset)                           &  \\
\texttt{control.jalr\_lsb}           & JALR target LSB clearing                                                       &  \\
\texttt{control.link\_reg}           & link register binding (rd $\gets$ PC+4 on JAL/JALR)                            &  \\
\bottomrule
\end{tabular}
\caption{Full obligation enumeration (1/2): Decode, ALU, Arithmetic, and Control groups.}
\label{tbl:full-obligations-a}
\end{table*}

\begin{table*}[!t]
\centering
\footnotesize
\setlength{\tabcolsep}{4pt}
\begin{tabular}{p{4.3cm}p{5.7cm}p{6.6cm}}
\toprule
\textbf{Obligation} & \textbf{Description} & \textbf{Detected benchmark bug} \\
\midrule
\multicolumn{3}{l}{\emph{Execution (7)}} \\
\midrule
\texttt{exec.operand\_bind}          & source/dest operand binding to register file                                   &  \\
\texttt{exec.opcode\_select}         & opcode selector exclusivity                                                    & \benchlist{OpenVM-Memory-Audit-o19\\Pico-ReadWrite-OpcodeSelector-01} \\
\texttt{exec.control\_flow}          & control-flow consistency between PC and executed instruction                   & Sp1-Pc-Audit-s28$^{\mathrm{A}}$ \\
\texttt{exec.mem\_effect}            & memory-effect binding to memory subsystem                                      &  \\
\texttt{exec.subword\_write}         & sub-word (byte/half) write boundary handling                                   &  \\
\texttt{exec.segment\_pow2}          & segment boundary $2^k$ threshold                                               & Risc0-ControlDone-Cycle-01 \\
\texttt{exec.bytecode\_bind}         & bytecode-to-operand binding (instruction word $\to$ decoded operands)          &  \\
\midrule
\multicolumn{3}{l}{\emph{Memory (12)}} \\
\midrule
\texttt{mem.addr\_align}             & address alignment for word/half accesses                                       &  \\
\texttt{mem.ts\_order}               & timestamp ordering (monotonic per address)                                     &  \\
\texttt{mem.addr\_route}             & address-space routing (RAM vs.\ I/O vs.\ private memory)                       & OpenVM-AddrSpace-Audit-o51 \\
\texttt{mem.sign\_ext}               & sign extension for LB/LH                                                       &  \\
\texttt{mem.store\_load\_flow}       & store-load value-flow consistency                                              & Nexus-Operand-01 \\
\texttt{mem.cap\_pow2}               & memory capacity $2^k$ threshold                                                & Nexus-MemorySize-01 \\
\texttt{mem.raw\_consistency}        & read-after-write consistency within a trace                                    &  \\
\texttt{mem.byte\_offset\_mask}      & byte offset masking within a word                                              &  \\
\texttt{mem.addr\_overflow}          & address overflow at $2^{32}$ boundary                                          &  \\
\texttt{mem.rw\_direction}           & read/write direction flag binding                                              & Sp1-Memory-Audit-s27 \\
\texttt{mem.init\_finalize}          & memory init/finalize boundary rows                                             &  \\
\texttt{mem.bounds}                  & memory range bounds enforcement                                                & OpenVM-RangeCheck-Audit-o25 \\
\midrule
\multicolumn{3}{l}{\emph{Lookup (5)}} \\
\midrule
\texttt{lookup.boolean\_mult}        & boolean multiplicity in lookup arguments                                       & OpenVM-Overflow-Audit-o1 \\
\texttt{lookup.xor\_mult}            & XOR multiplicity in lookup arguments                                           &  \\
\texttt{lookup.acc\_pow2}            & lookup accumulator $2^k$ threshold                                             &  \\
\texttt{lookup.range\_pow2}          & range-table $2^k$ threshold                                                    &  \\
\texttt{lookup.table\_pop}           & lookup table population (every used row is materialized)                       &  \\
\midrule
\multicolumn{3}{l}{\emph{Interaction (2)}} \\
\midrule
\texttt{interaction.digest}          & cross-component digest routing                                                 &  \\
\texttt{interaction.commit\_pow2}    & commitment vector $2^k$ threshold                                              & Jolt-Dory-ShortTrace-01$^{\mathrm{A}}$ \\
\midrule
\multicolumn{3}{l}{\emph{Row (2)}} \\
\midrule
\texttt{row.padding\_leak}           & padding row interaction leakage into live constraints                          & OpenVM-Timestamp-Audit-o3 \\
\texttt{row.trace\_pow2}             & trace padding $2^k$ threshold                                                  & Jolt-Dory-ShortTrace-01$^{\mathrm{A}}$ \\
\midrule
\multicolumn{3}{l}{\emph{Time (3)}} \\
\midrule
\texttt{time.boundary\_origin}       & timestamp boundary origin consistency                                          & \benchlist{OpenVM-Timestamp-Audit-o2\\OpenVM-Timestamp-Audit-o26} \\
\texttt{time.diff\_range}            & monotonic timestamp difference range                                           &  \\
\texttt{time.fp\_wrap}               & $\mathbb{F}_p$ modular wrap on timestamp arithmetic                            &  \\
\bottomrule
\end{tabular}
\caption{Full obligation enumeration (2/2): Execution, Memory, Lookup, Interaction, Row, and Time groups.}
\label{tbl:full-obligations-b}
\end{table*}

\end{document}